\documentclass[showpacs,prb,twocolumn]{revtex4}

\usepackage[english]{babel}
\usepackage{amssymb,amsmath}
\usepackage{color}
\usepackage{epsfig}
\usepackage{graphicx}
\usepackage{ulem}

\usepackage{CJK}

\begin{document}
\begin{CJK*}{Bg5}{bsmi}

\title{Edge state effects in junctions with graphene electrodes}

\author{Dmitry\,A.~Ryndyk\footnote{New affiliation: Institute for Materials Science, Dresden University of Technology, D-01069 Dresden, Germany}, Jan~Bundesmann, Ming-Hao~Liu (¼B©ú»¨), and Klaus~Richter}

\affiliation{Institute for Theoretical Physics, University of Regensburg, D-93040 Regensburg, Germany}

\date{\today}

\begin{abstract}
We consider plane junctions with graphene electrodes, which are formed by a single-level system (``molecule'') placed between the edges of two single-layer graphene half planes. We calculate the edge Green functions of the electrodes and the corresponding lead self-energies for the molecular levels in the cases of semi-infinite single-layer electrodes with armchair and zigzag edges. We show two main effects: first, a peculiar energy-dependent level broadening, reflecting at low energies the linear energy dependence of the bulk density of states in graphene, and, second, the shift and splitting of the molecular level energy, especially pronounced in the case of the zigzag edges due to the influence of the edge states. These effects give rise to peculiar conductance features at finite bias and gate voltages.   
\end{abstract}

\pacs{\vspace{-0.1cm} 73.63.-b, 85.65.+h, 72.80.Vp}

\maketitle
\end{CJK*}

\section{Introduction}

Single-molecule nanosystems are in the focus of experimental and theoretical investigations in recent years. One branch of the basic research in this field is concentrating on the question of the current through molecular junctions with metallic or semiconductor electrodes. Many phenomena, typical for nanoscale transport, such as Coulomb blockade, Kondo effect, vibronic and polaronic effects, to name a few, have been observed and explained. A comprehensive bibliography  can be found in Refs.\,~\cite{Cuniberti05book,Cuevas10book,Song11advmat}. However, despite the experimental progress and the theoretical efforts, the understanding of the properties of single molecules coupled to metal electrodes, especially their transport properties, is far from being satisfactory. One of the issues is still the often poor reproducibility of the experimental results. Unfortunately, the structure and quality of molecular junctions are not completely controlled yet. One of the main problems is the size mismatch between metal electrodes and molecules and the impossibility to control the metal-to-molecule interface at nanoscale. Besides, gold, the most popular electrode material, has high atomic mobility and at room temperature the  geometry of nanoelectrodes is not completely stable. Thus, traditional electrodes show their principle limits and other materials for molecular electronics should be considered.

Carbon based materials, e.g., fullerene, nanotube or graphene, have the advantage of well controlled crystal structures, stability up to high temperatures due to $sp2$ covalent bonds, and appropriate sizes. Production of ultrathin epitaxial graphite films~\cite{Berger04jpcb} opens new ways for electrode technology. 

Although the control of the atomic structure of graphene edges is also a challenge for present-day lithography, important progress has recently been achieved~\cite{Jia09science,Jin09prl,Chuvilin09njp,He10apl,Song11nanolett}. It is also important that carbon electrodes allow for many ways to anchor organic and inorganic molecules, thus being promising for functional devices. Several types of carbon nanoelectrodes were suggested  with different geometry and dimensionality. One can mention, as an example, the study of electron transport across molecular junctions with carbon nanotube electrodes (see, e.g., Refs.~\onlinecite{Gutierrez02prb,Chen07prb} and references therein).

\begin{figure}[b]
\begin{center}
\epsfxsize=0.8\hsize
\epsfbox{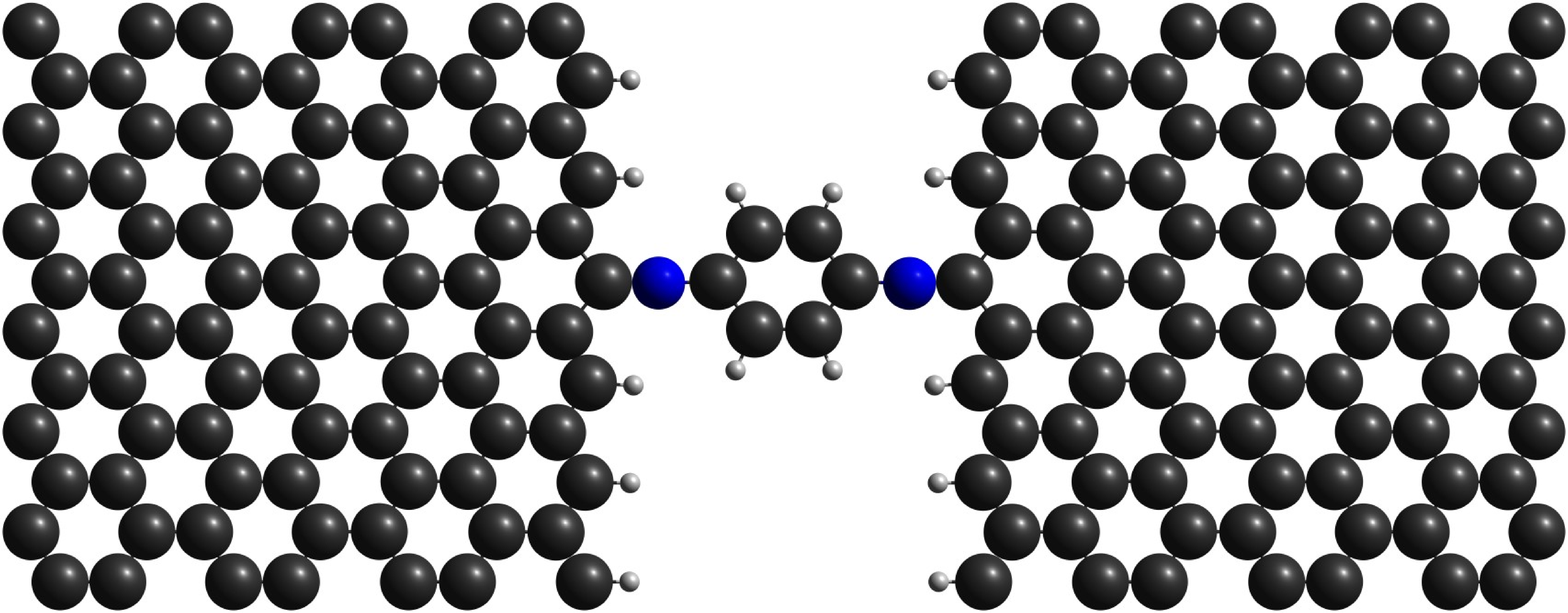}
\epsfxsize=1.\hsize
\vskip 0.2cm
\epsfbox{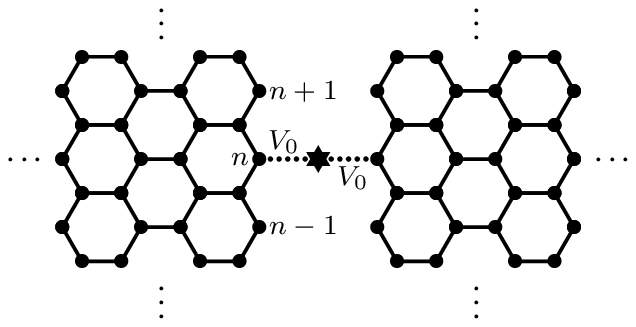}
\caption{(Color online) Example of a plane graphene molecular junction (upper panel) and its schematic representation (lower panel) of the considered single-level model for zigzag edge.}
\label{CNE_Graphene}
\end{center}
\end{figure}

\vskip 0.4cm
The advantage of plane graphene (single- or multilayer) is that graphene sheets are suitable for nanoscale lithography~\cite{Berger04jpcb,Jia09science,Jin09prl,Chuvilin09njp,He10apl,Song11nanolett} and can be considered as a base for molecular integrated circuits. Thus we focus on molecular junctions with plane graphene electrodes (Fig.\,\ref{CNE_Graphene}), where a molecule is coupled to the graphene edges (here a variety of molecules and anchoring groups can be used). The theoretical investigation of such structures was started recently for all-carbon junctions~\cite{Fagas04cpl,Fagas0403694,Cheraghchi08prb,Yin09jcp,Chen09prb,Fuerstl10epl,Shen10jacs,Saha10prl,Kawai11prb} and junctions with different organic molecules.~\cite{Agapito07jpcc,Motta11prb,Cai11eprint,Carrascal12eprint} In particular, one can mention the experimental~\cite{Jin09prl,Chuvilin09njp} and theoretical~\cite{Cheraghchi08prb,Chen09prb,Jin09prl,Fuerstl10epl,Shen10jacs} investigations of linear atomic carbon chains between single-layer graphene electrodes and first experimental observation of the  current through the molecular junction with few-layer electrodes.~\cite{Prins11nanolett} 

The other advantage of graphene electrodes is that the molecular gating problem could be solved. Indeed, large metal electrodes screen the external gate potential and makes it almost impossible to shift molecular levels in a controlled way. Oppositely, in plane structures with thin graphene electrodes the gate potential can be used rather efficiently.

The simplest possible case of a molecular bridge with graphene electrodes is a junction formed in a nanogap with armchair or zigzag edges. The zigzag edge case is shown schematically in Fig.\,\ref{CNE_Graphene}. The armchair edge is obtained if one cuts a graphene sheet in the perpendicular direction. The transport properties of such junctions are determined by the peculiarities of the edge Green functions of semi-infinite graphene electrodes. In the case of zigzag edge the main features are influenced by edge states.  In this paper we address the question of how these properties of graphene electrodes affect charge transport through the molecular bridge. Hence we do not focus on molecular specific properties and replace the molecule by one spin-degenerate noninteracting level. The electrodes are described in the framework of the \mbox{$\pi$-electron} tight-binding model.~\cite{Wallace47prb} This approach is complementary to an ab initio one and allows to obtain physically transparent results.

Below we consider the tight-binding model of graphene molecular junctions with armchair and zigzag edges (Sec.\,\ref{sec:sl_model}). We calculate the edge Green functions and lead self-energies in Sec.\,\ref{sec:egf}. Then, in Sec.\,\ref{sec:levels} the edge dependent level shift, splitting and broadening are considered. Finally, in Sec.\,\ref{sec:current} we discuss the current at finite voltage and the gating effect, and give conclusions in Sec.\,\ref{sec:conclusions}.

\section{The tight-binding  model}
\label{sec:sl_model}

We write the tight-binding Hamiltonian of our system (for fixed spin $\sigma=\uparrow$ or $\sigma=\downarrow$, the spin index is omitted) as
\begin{align}\nonumber
   \hat H= & (\epsilon_0+eV_g+e\varphi_0)d^\dag d+\sum_{s=L,R}\left(V^*_0c^{\dag}_{sn}d+V_0d^{\dag}c_{sn}\right) \\ \label{H}
& +\sum_{s;\,i} (eV_{gl}+e\varphi_s)\,c^{\dag}_{si}c_{si} +\sum_{s;\,ij} t_{ij}\,c^{\dag}_{si}c_{sj}.
\end{align}
Here $d^\dag$, $d$ are the creation and annihilation operators for a molecular level, while the $c_{si}^\dag$, $c_{si}$ operators describe the lattice tight-binding local orbital at the $i$-th site in the $s$-th (left or right) electrode. The molecular level is assumed to be coupled only to one edge site of the graphene lead with site index $n$ (Fig.\,\ref{CNE_Graphene}) and $V_0$ is the coupling matrix element. We denote the tight-binding hopping matrix elements $t_{ij}$ between lattice sites by $t_{ij}=t$ for nearest neighbor sites, and $t_{ij}=t'$ for next-nearest neighbor sites of the lattice. $V_g$ is the gate voltage acting on the molecular level and $V_{gl}$ is the gate voltage, applied to the leads. At finite bias voltage $V$ (defined by the left and right electrical potentials, \mbox{$V=\varphi_L-\varphi_R$}) the energy of the molecular level is shifted. In the linear approximation this shift is described by a parameter $\eta$: $\varphi_0=\varphi_R+\eta(\varphi_L-\varphi_R)$,
where $0 < \eta < 1$ characterizes the symmetry of the voltage drop across the junction, and $\eta=0.5$ stands for the symmetric case (when it is convenient to use \mbox{$\varphi_L=V/2$} and \mbox{$\varphi_R=-V/2$}). The generalization of this model to the case of many molecular levels, as well as interacting molecular levels, is straightforward.

Note that while the lead gate potential $V_{gl}$ and the bias potentials $\varphi_s$ enter the Hamiltonian (\ref{H}) on an equal footing, the physical sense and the effect of these two potentials are different. The gate potential shifts the energy levels in the finite-size region of the electrodes near the molecule, where it is applied, but does not change the Fermi level, which is determined by the large equilibrium electrodes outside this region. The bias potential shifts additionally the energy distribution of electrons in the electrodes according to
\begin{equation}\label{fs}
f_s^0(\epsilon)=\frac{1}{\exp\left((\epsilon-e\varphi_s)/k_BT\right)+1}
\end{equation}
with temperature $T$.

The gate potentials $V_{gl}$ and $V_{g}$ of the leads and the molecule can generally differ, but in this paper we consider the case $V_{g}=V_{gl}$ being a good approximation for plane structures.

\section{Edge Green functions and self-energies}
\label{sec:egf}

There are two electronic properties that determine the main peculiarities of graphene edges as electrodes. First, the energy dependence of the local density of states in bulk graphene $\rho(\epsilon)$ has a minimum near the Fermi energy of undoped graphene (to be chosen as the zero energy) and is almost linear near this point $\rho(\epsilon)\propto|\epsilon|$. The coupling of molecular electronic levels to the electrodes is characterized by the lead self-energy $\Sigma(\epsilon)$, which is proportional to the edge Green function of the electrode $G_{nm}(\epsilon)$. If a single molecular level is coupled to one edge atom of the electrodes, as in our case, the retarded (lesser) self-energies are
\begin{equation}\label{SigmaR}
\Sigma_s^{R(<)}(\epsilon)=\left|V_0\right|^2{G}^{R(<)}_{s,nn}(\epsilon),
\end{equation}
where ${G}^{R(<)}_{s,nn}$ is the Green function of the $s$-th lead at the site $n$, to which the molecular level is connected.

\begin{figure}[t]
\begin{center}
\epsfxsize=1.\hsize
\epsfbox{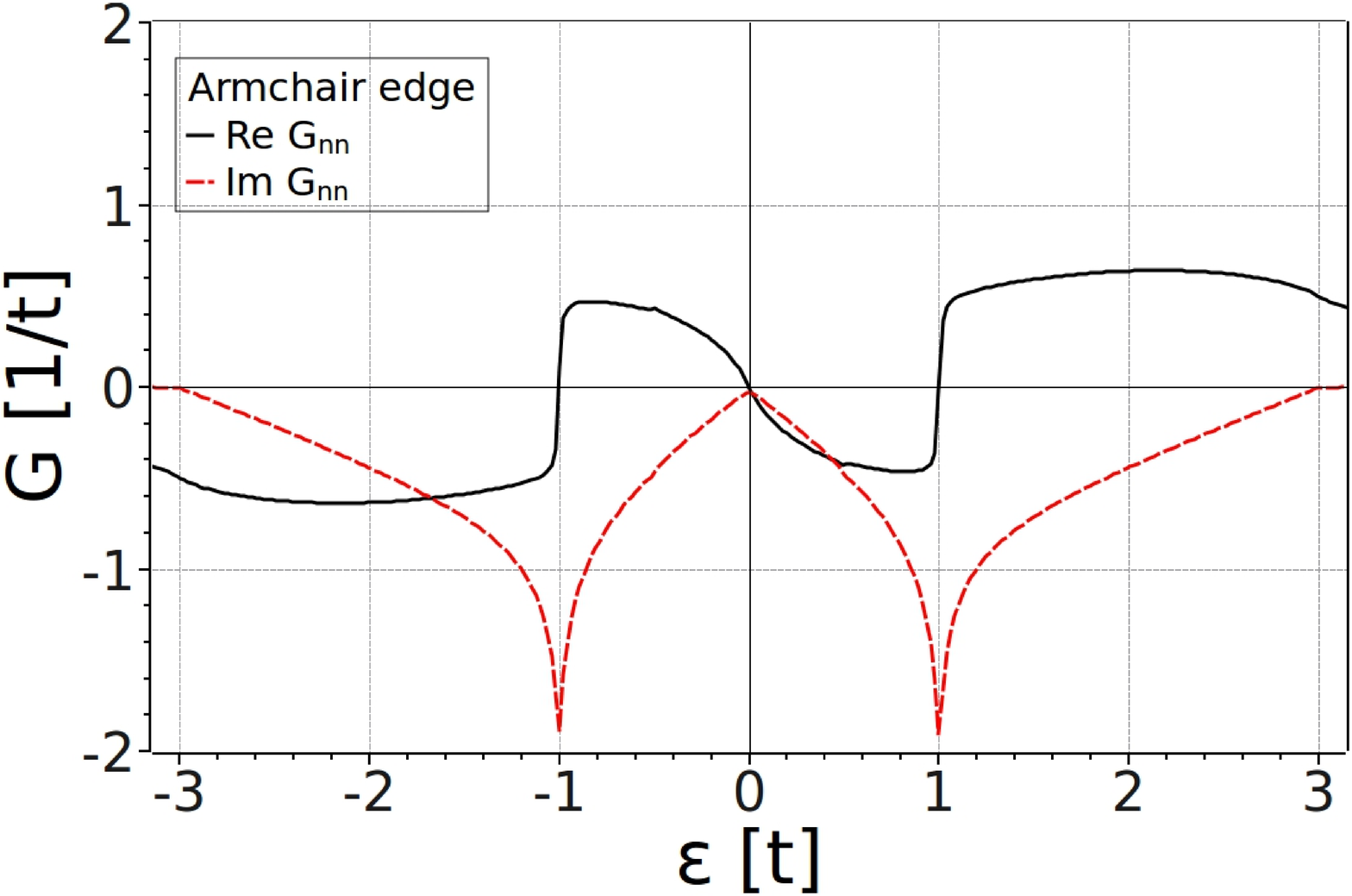}
\caption{(Color online) Edge Green function $G_{nn}(\epsilon)$ for the armchair edge.}
\label{EGF_ac}
\end{center}
\end{figure}

\begin{figure}[b]
\begin{center}
\epsfxsize=1.\hsize
\epsfbox{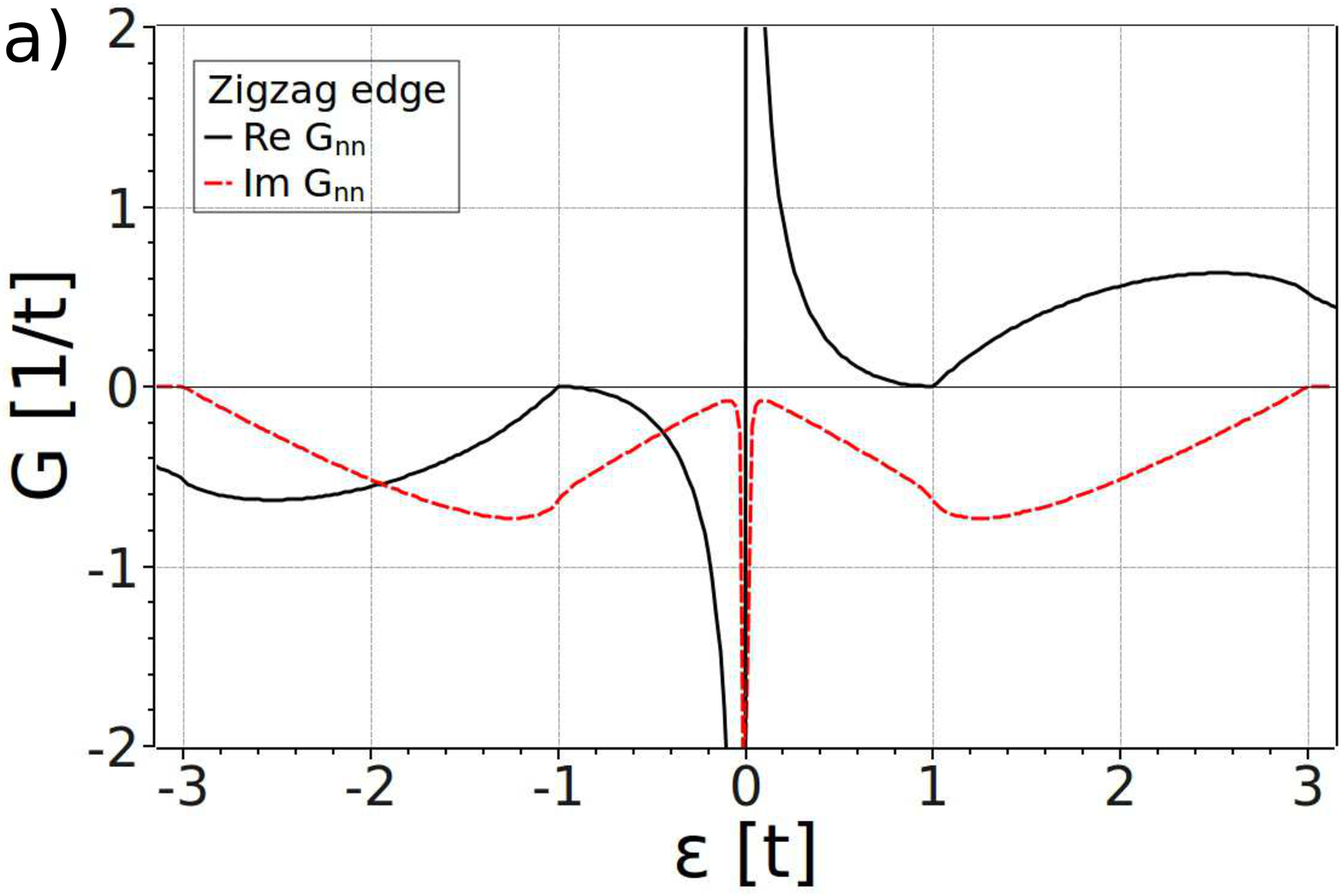} \\
\epsfxsize=1.\hsize
\epsfbox{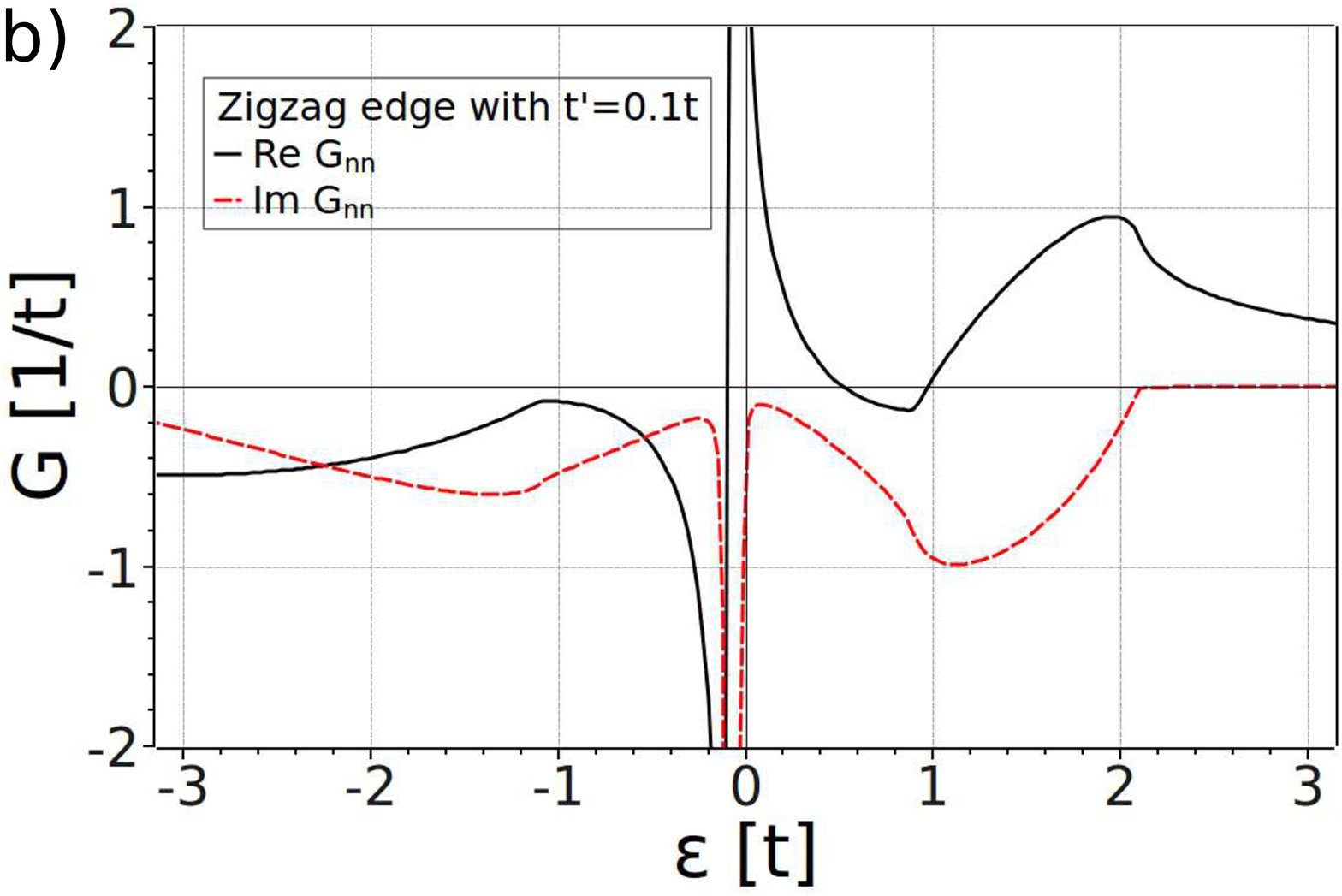}
\caption{(Color online) Edge Green function $G_{nn}(\epsilon)$ for the zigzag edge with (a) only nearest neighbor hopping $t$ and (b) additional next-nearest neighbor hopping $t'=0.1t$.}
\label{EGF_zz}
\end{center}
\end{figure}

The self-energy is a complex function, with the imaginary part known as the level-width function \mbox{$\Gamma(\epsilon)\propto -{\rm Im}\Sigma^R(\epsilon)\propto|V_0|^2\rho(\epsilon)$} (actually \mbox{$\rho_n(\epsilon)=-\pi{\rm Im}G^R_{nn}(\epsilon)$} at lattice site $n$). The level-width function determines, in particular, the broadening of molecular levels and the maximal (resonant) current through these levels. Thus, the current will strongly depend on the relative position of the molecular level, the Fermi level and the Dirac point (the energy of minimal $\rho(\epsilon)$). This is true for both armchair and zigzag edges. Moreover the zigzag edge supports so-called edge states,~\cite{Nakada96prb,Fujita96jpsj,Niimi06prb,Kobayashi06prb} which are localized near the edge with energies near the Dirac point.

We performed numerical calculations of the edge Green functions and lead self-energies (\ref{SigmaR}) for armchair and zigzag graphene edges using the iterative method~\cite{LopezSancho84jpf,LopezSancho85jpf} and the eigendecomposition based method.~\cite{Wimmer08thesis} The results for the edge retarded Green function are presented in Figs. \ref{EGF_ac} and \ref{EGF_zz}. We show here only the diagonal part $G^R_{nn}$ of the full matrix Green function, required to calculate the self-energy. We checked the influence of additional next-nearest neighbor coupling $t'$ (Fig.\,\ref{EGF_zz}b) and found that the main presented results are qualitatively the same. Thus we focus on the case $t'=0$ in this paper.   

In both, of the armchair and zigzag cases we found the linear energy dependence of the imaginary part (density of states) at small energies, which reflects the bulk properties of graphene. In the zigzag case an additional feature in the density of states is present near zero energy, and the real part of the self-energy is quite different from the armchair case. Because of the weak energy dependence on the quasi-momentum (along the edge) the edge states form a flat energy band with a $\delta$-function type density of states giving rise to the divergence close to $\epsilon=0$. Correspondingly, the real parts of $G_{nn}(\epsilon)$ at the edge and $\Sigma(\epsilon)$ have a singularity at this energy (Fig.\,\ref{EGF_zz}). Since the real part of the self-energy renormalizes the energy of the molecular levels, $\epsilon'\approx\epsilon_0+{\rm Re}\Sigma(\epsilon_0)$, the energies of levels at low energies are strongly shifted. Moreover, the spectral function, and hence the dressed energy levels, can be split because the sign of ${\rm Re}\Sigma(\epsilon)$ changes close to the singularity point.

\section{Level shift, splitting and broadening}
\label{sec:levels}

Using the calculated self-energies we investigated the spectral function
\begin{equation}\label{eq-A}
A(\epsilon)=-2{\rm Im}G^R(\epsilon)
\end{equation}
of a single level, coupled to armchair and zigzag electrodes. The retarded Green function of the level is
\begin{equation}\label{GR}
  G^R(\epsilon)=\frac{1}{\epsilon-\epsilon_0-\Sigma^R_L(\epsilon)-\Sigma^R_R(\epsilon)},
\end{equation}
where the lead self-energies $\Sigma^R_L(\epsilon)$ and $\Sigma^R_R(\epsilon)$ are calculated from the Green functions of the leads using expression (\ref{SigmaR}). 

The renormalized resonant level position $\epsilon'$ is determined from the maximum of the spectral function, Eqs.\,(\ref{eq-A},\ref{GR}):
\begin{equation}\label{level}
  \epsilon'-\epsilon_0-{\rm Re}\left[\Sigma^R_L(\epsilon')+\Sigma^R_R(\epsilon')\right]=0.
\end{equation}
For the armchair case (Fig.\,\ref{EGF_ac}) the graphical solution of this equation has one solution in most cases, but three solutions at $\epsilon_0\sim \pm t$. In the zigzag case (Fig.\,\ref{EGF_zz}) there are always two solutions because of the sign-changing singularity at small energies. The existence of two solutions means that the original level is split into two sublevels. This splitting can be understood as a result of the hybridization of the molecular level with the edge states. The magnitude of the  splitting (the distance between the levels), as well as the shift and broadening, are controlled by the coupling $V_0$ of the molecular level to the leads. Below we consider mainly the most interesting case $V_0\ll t$ and choose $V_0\sim 0.1t$, which is typical for covalent coupling of organic molecules to the leads.

The broadening of the resonant level with energy $\epsilon'$ is determined by the imaginary parts of the self-energies,
\begin{equation}
  \Gamma=-2{\rm Im}\left[\Sigma^R_L(\epsilon')+\Sigma^R_R(\epsilon')\right].
\end{equation}
It is mainly given by the graphene bulk density of states (with some quantitative variations in the armchair and zigzag cases). Additionally there is a sharp peak of the density of states at the energy of the edge state; however, the molecular levels are shifted from these energies.

\begin{figure}[b]
\begin{center}
\epsfxsize=0.8\hsize
\epsfbox{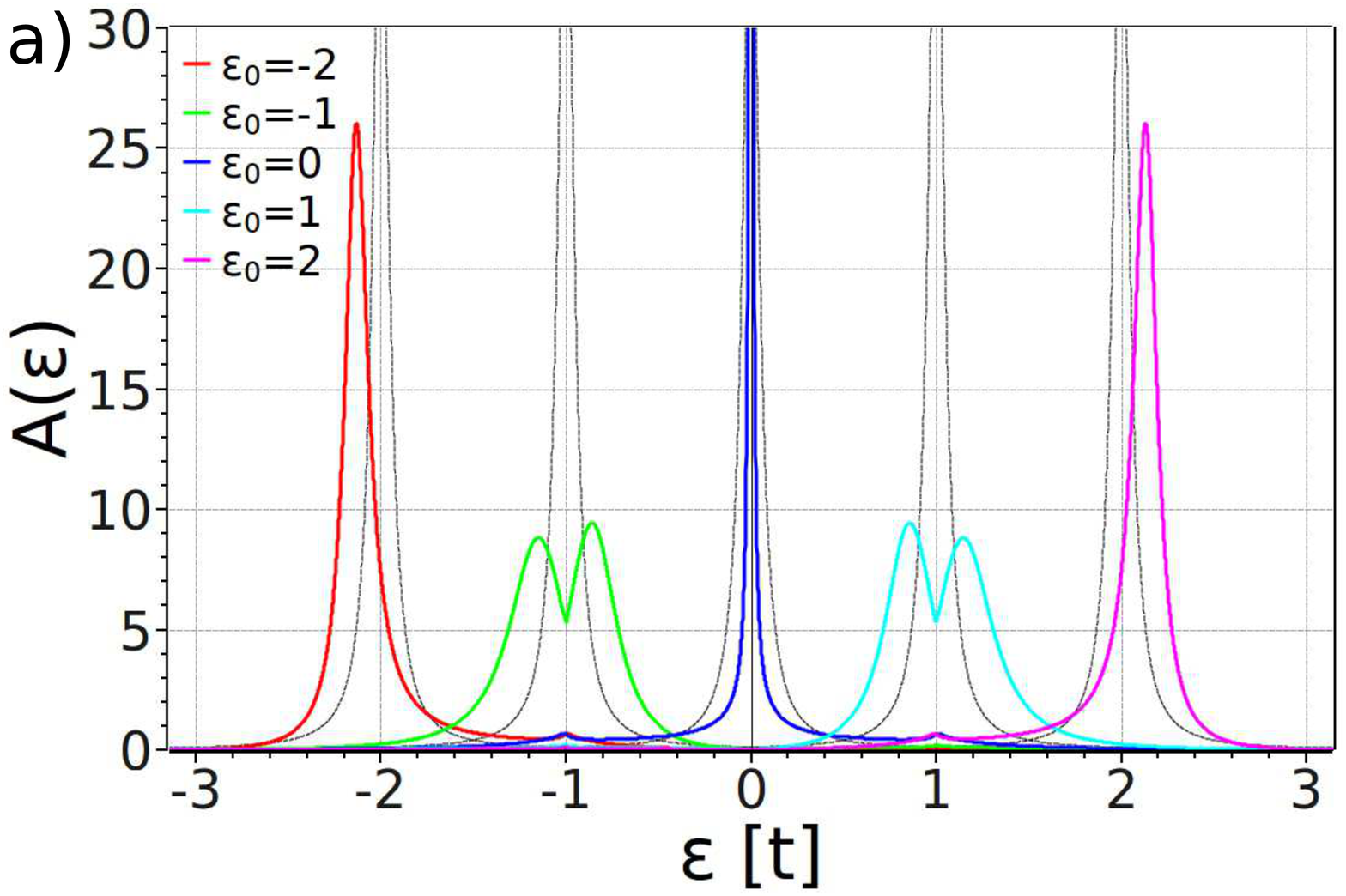}
\epsfxsize=0.8\hsize
\epsfbox{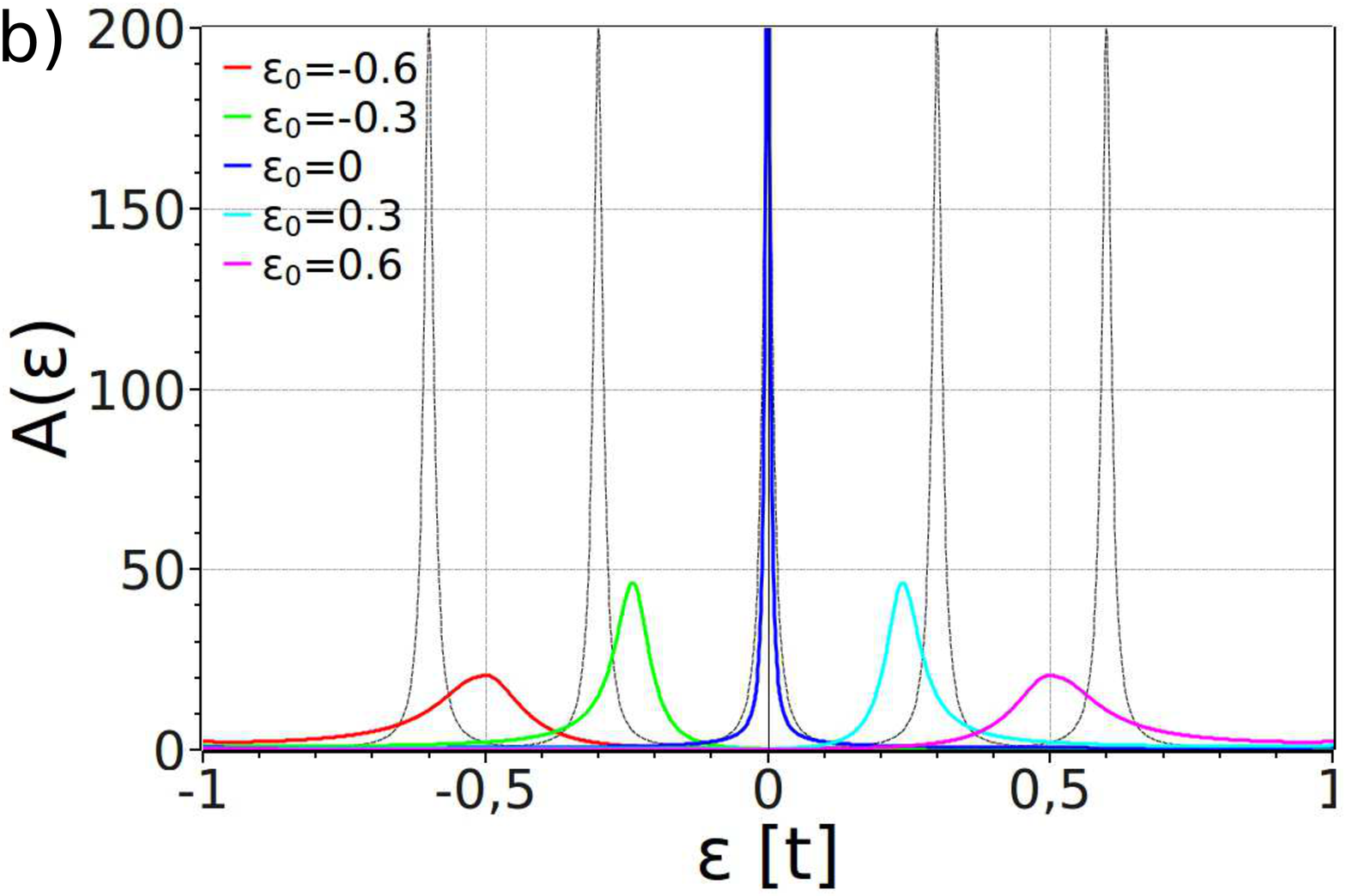}
\caption{(Color online) Spectral function of a single level coupled to armchair edges. The equidistant dashed lines show the position of the level in the wide-band limit with $V_0=0.1t$.}
\label{SL_ac}
\end{center}
\end{figure}

\begin{figure}[b]
\begin{center}
\epsfxsize=0.81\hsize
\epsfbox{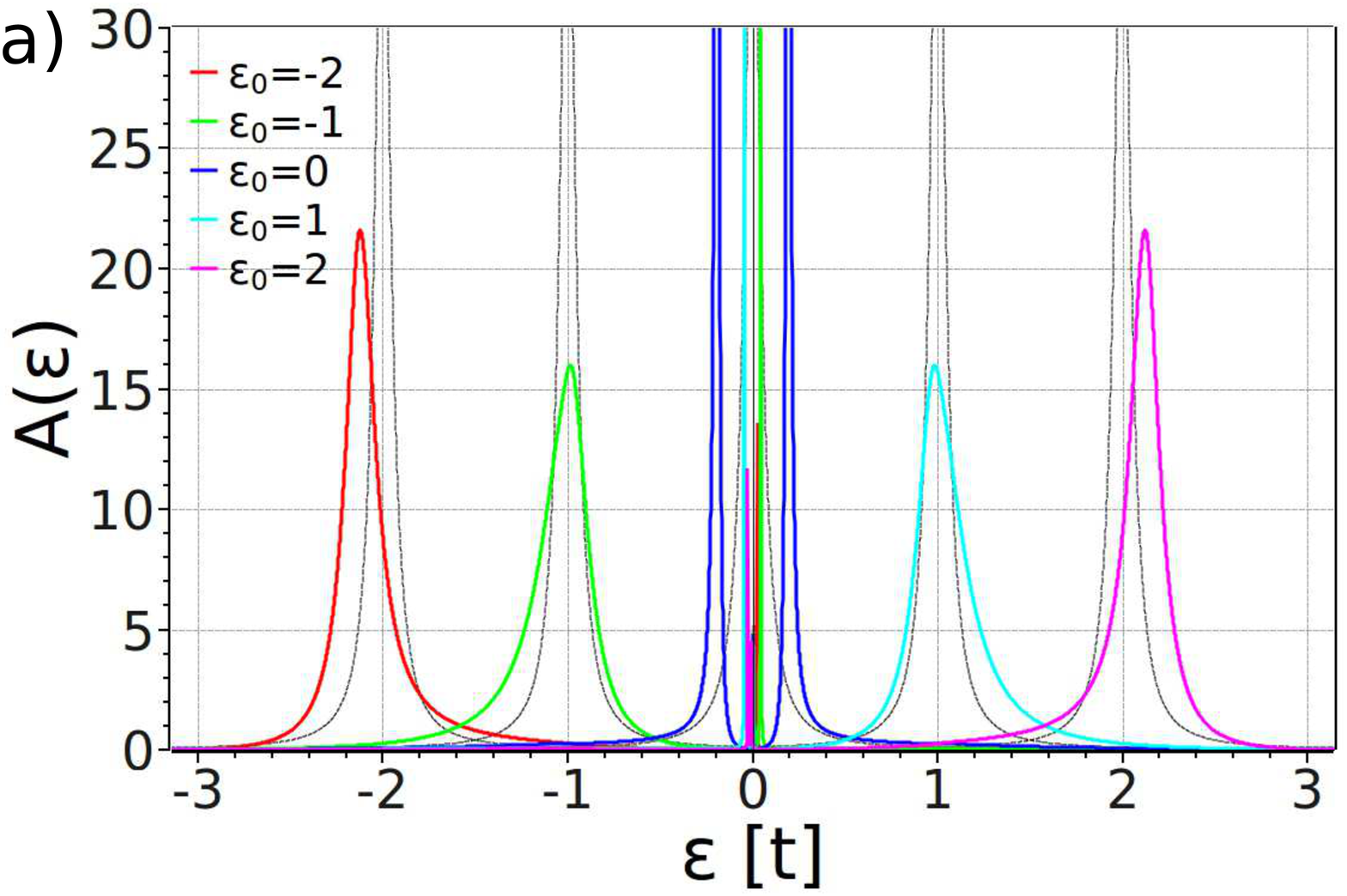}
\epsfxsize=0.81\hsize
\epsfbox{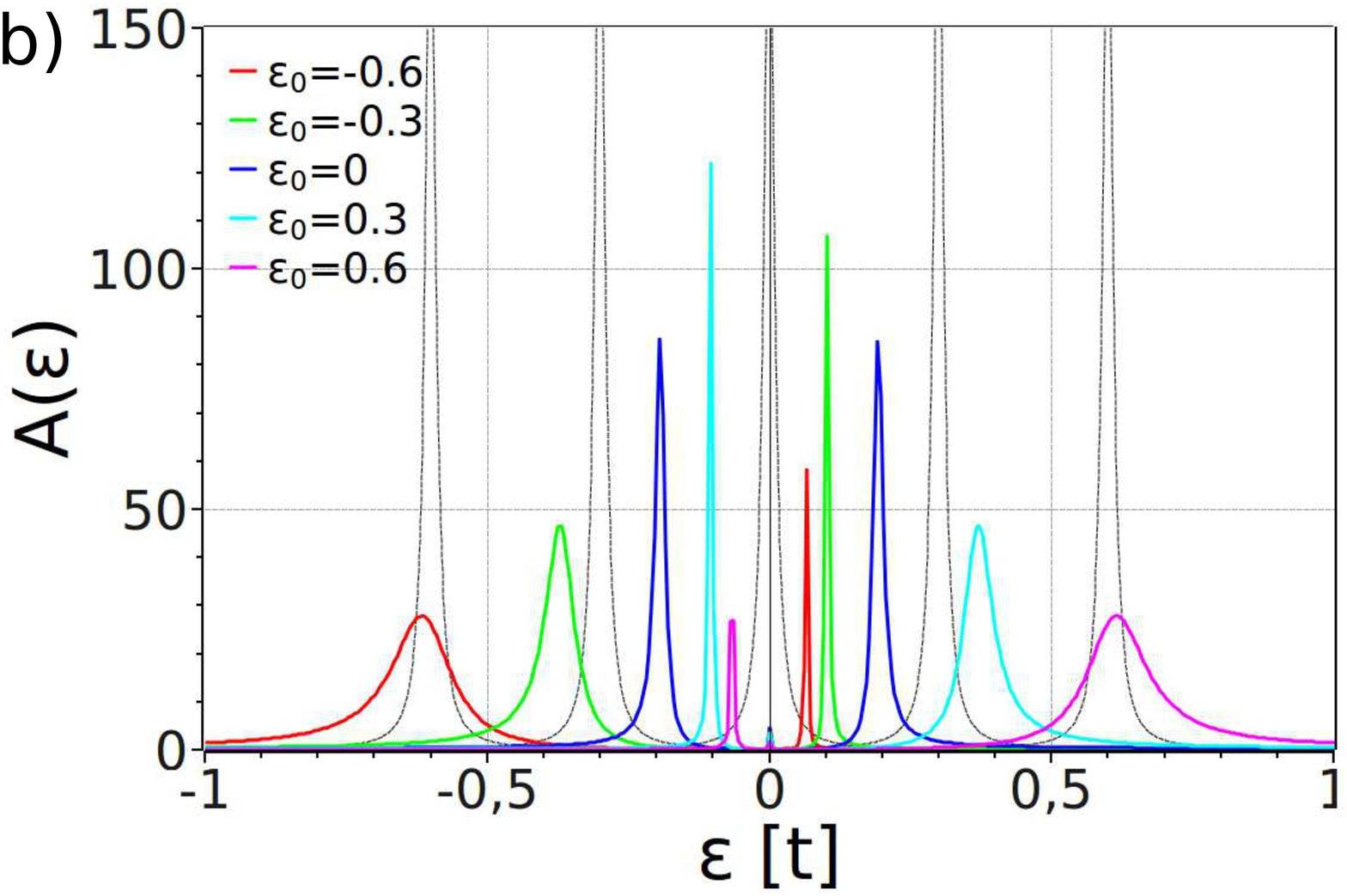}
\caption{(Color online) Spectral function of a single level coupled to zigzag edges. The dashed lines show the position of the levels in the wide-band limit with $V_0=0.1t$.}
\label{SL_zz}
\end{center}
\end{figure}

Figures \ref{SL_ac} and \ref{SL_zz} show $A(\epsilon)$ for armchair and zigzag lead termination, respectively, as a function of the energy for different original level positions $\epsilon_0$. For comparison we present also the curves in the wide-band limit (\mbox{$\Sigma^R_s(\epsilon)=-(i/2)\Gamma_s=const$}), commonly chosen for metal leads. As we already anticipated, the main results are a energy level position dependent broadening for both edge types and a level shift and splitting mainly for the zigzag edge. Of course, the results also depend on the coupling $V_0$.


In the armchair case (Fig.\,\ref{SL_ac}) the main effect on $A(\epsilon)$ at small level energies ($\epsilon_0<t$) is the energy-dependent level broadening due to the imaginary part of the self-energy, but the level shift due to the real part of the self-energy is also clearly observed (Fig.\,\ref{SL_ac}b). The peak at $\epsilon_0=0$ is very sharp, because the imaginary part vanishes at $\epsilon=0$. At larger level energies the spectral function has a more complex form (Fig.\,\ref{SL_ac}a). The levels at $\epsilon_0\approx\pm t$ are split, because ${\rm Re \Sigma^R}$ changes sign at $\epsilon=\pm t$ (Fig.\,\ref{EGF_ac}). However due to the maximal level broadening at these energies, the splitting is not clearly pronounced and the levels overlap. At larger energies the broadening and shift decreases.


In the zigzag case (Fig.\,\ref{SL_zz}) the singularity of ${\rm Re \Sigma^R}$ at zero energy changes the picture drastically, especially for small level energies. First of all, the spectral density is ``repelled'' from small energies. If the level is originally at $\epsilon_0=0$, it is split into two sublevels. At large $|\epsilon_0|$ the small part of the integral spectral density of the original level is split and the second sublevel appears at the other side of the $\epsilon=0$ point. For large energies of the original level, the second sublevel is close to zero energy (position of the edge state), but its spectral weight is small.


\begin{figure}[t]
\begin{center}
\epsfxsize=0.95\hsize
\epsfbox{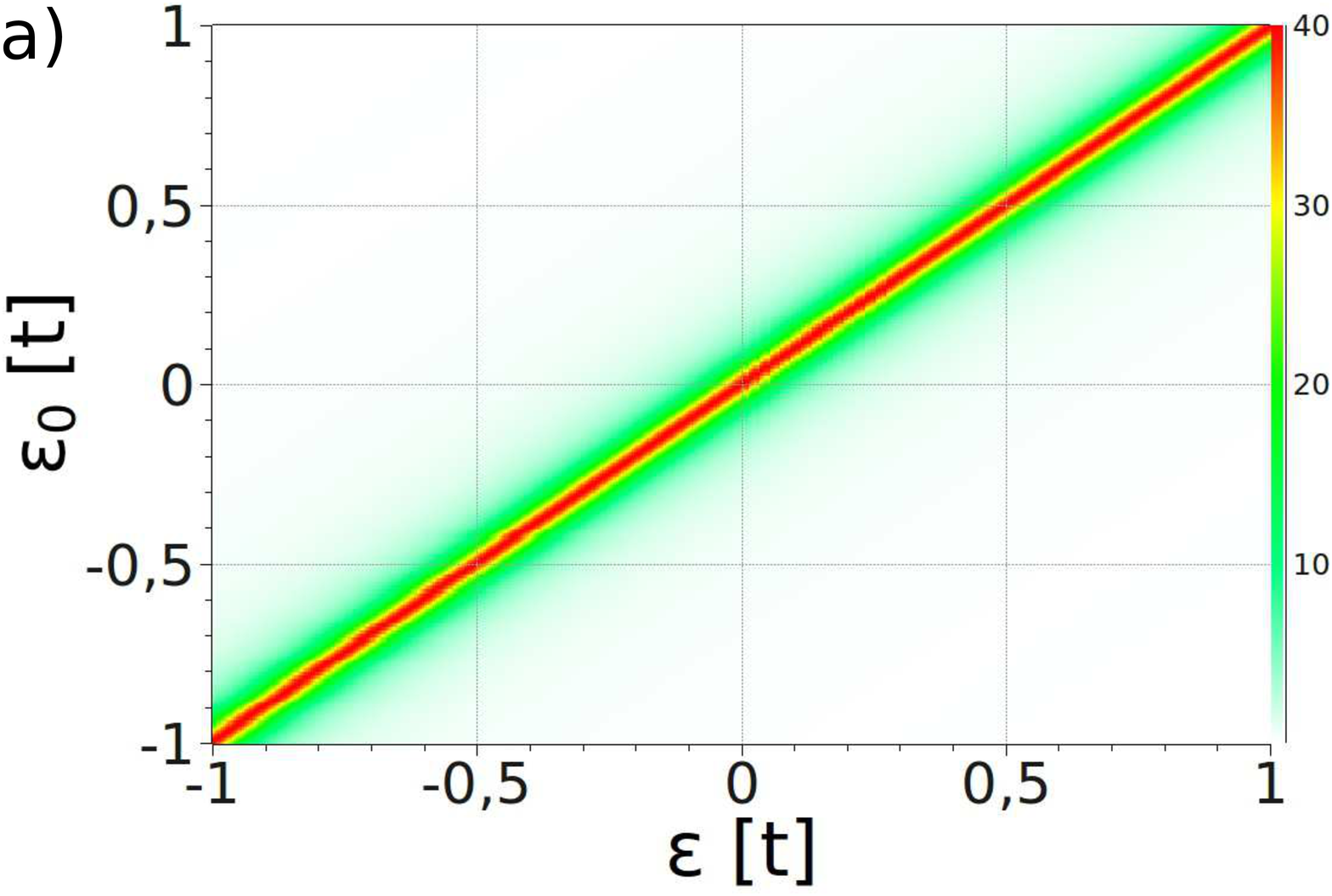}
\epsfxsize=0.95\hsize
\epsfbox{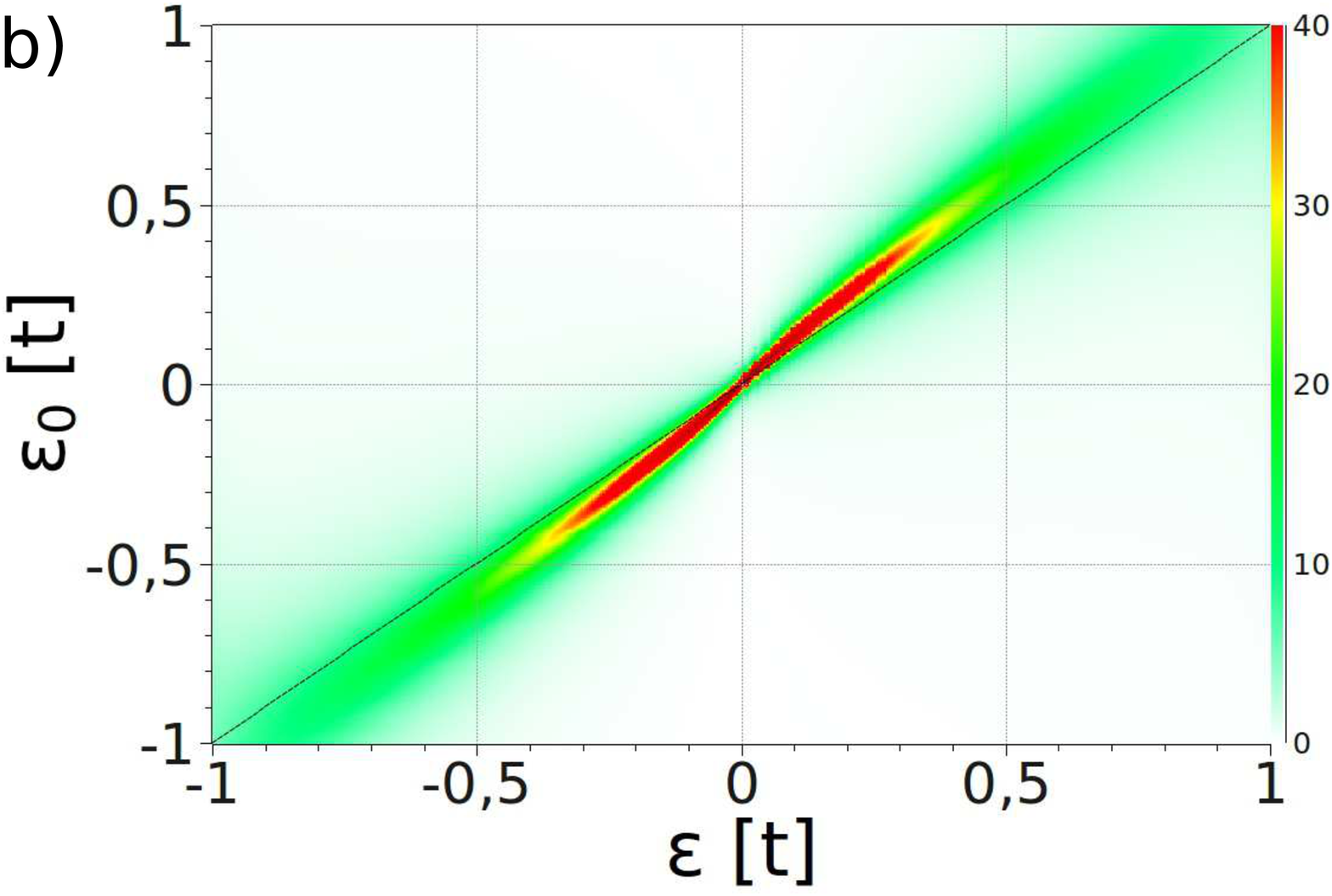}
\epsfxsize=0.95\hsize
\epsfbox{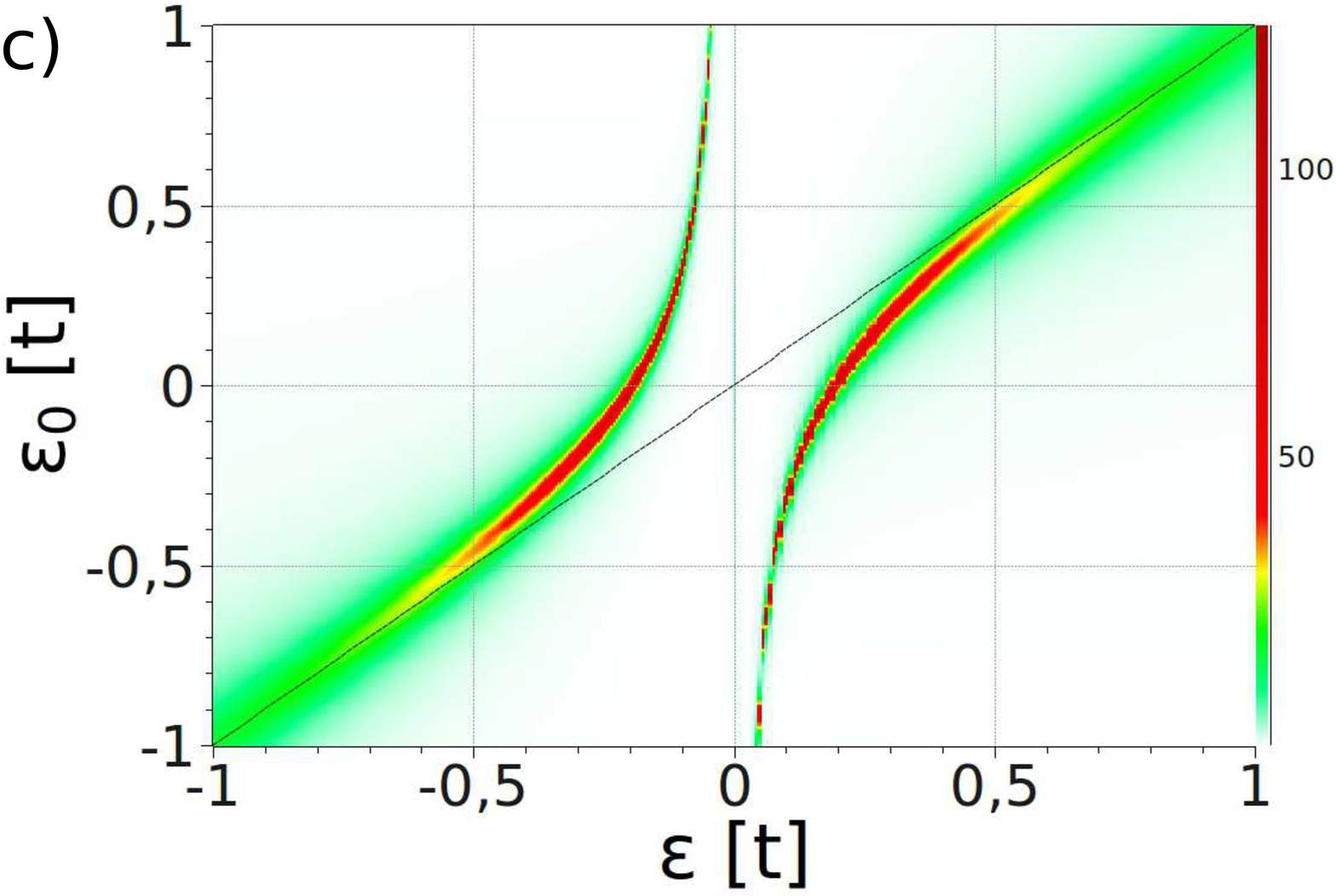}
\caption{(Color online) Color-coded strength of spectral function of a single level coupled to wide-band leads (a), armchair leads (b) and zigzag leads (c). The scale of the spectral function is shown at the right.}
\label{SL_Ac}
\end{center}
\end{figure}

We summarize these findings in the 2D plot, Fig.\,\ref{SL_Ac}. Here the spectral function of a single level is shown color-coded as a function of the energy $\epsilon$ and the level position $\epsilon_0$ in three cases: the wide-band limit (for comparison), the armchair leads and the zigzag leads. All the three effects: level broadening, shift and splitting are visible in Fig.\,\ref{SL_Ac}(b) and Fig.\,\ref{SL_Ac}(c). The strong energy dependence of the edge Green functions and correspondingly of the lead self-energies of graphene electrodes leads to the qualitatively different behavior of the molecular levels, coupled to such electrodes, compared to the case of wide-band electrodes. The shift of the level position and the broadening depend strongly on the energy of the unperturbed molecular level $\epsilon_0$. Besides, in the zigzag case, the edge states of the leads are hybridized with the molecular level, and the molecular level is split into two sublevels.

Such unusual interplay of level shift and broadening is expected to result in peculiar transport properties through graphene molecular junctions, compared to the case of metal wide-band electrodes.

\section{Current and differential conductance: gating and edge state effects}
\label{sec:current}


Finally we calculate the current and differential conductance with the assumption that graphene electrodes (including the edge states) are kept in  equilibrium, but at different electrical potentials due to the bias voltage $V$. This condition can be easily fulfilled if coupling of a molecule to the leads is weak enough. We follow the formulation pioneered by Meir, Wingreen and Jauho.~\cite{Meir92prl,Jauho94prb,Jauho06jpcs} The current from the left ($s=L$) or right ($s=R$) lead into the central system is described by the  expression (here we consider the case of spin-unpolarized leads)
\begin{align}\label{J}\nonumber
 J_{s}=\frac{\mathrm{i}e}{\hbar}\int\frac{d\epsilon}{2\pi}{\rm Tr} & \left\{
 {\Gamma}_s(\epsilon)\left({G}^<(\epsilon)+ \right.\right.\\
& \displaystyle \left.\left.
 + f^0_s(\epsilon)
 \left[{G}^R(\epsilon)-{G}^A(\epsilon)\right]\right)\right\}.
\end{align}
Here $f^0_s(\epsilon)$ is the equilibrium Fermi distribution function in the $s$-th lead, Eq.\,(\ref{fs}), $\Gamma_s(\epsilon)=-2{\rm Im}\Sigma^R_s(\epsilon)$ is the level-width function, ${G}^R(\epsilon)$ is the retarded function of the level, as defined by Eq.\,(\ref{GR}), ${G}^A(\epsilon)=\left[{G}^R(\epsilon)\right]^*$, and ${G}^<(\epsilon)$ is the lesser Green function. It can be found from the Dyson-Keldysh equation in the integral form,
\begin{equation}\label{GR_Int}
{G}^<(\epsilon)={G}^R(\epsilon)\left({\Sigma}_L^<(\epsilon)+{\Sigma}_R^<(\epsilon)\right){G}^A(\epsilon),
\end{equation}
where the lesser self-energy of the noninteracting leads is
\begin{equation}
{\Sigma}^<_s(\epsilon)=i{\Gamma}_s(\epsilon)f^0_s(\epsilon).
\end{equation}

The current is a function of the bias voltage \mbox{$V=\varphi_L-\varphi_R$}. The second parameter is the gate voltage $V_{gl}$ which is assumed to shift both the energy level $\tilde\epsilon=\epsilon_0+eV_{gl}$ of the molecule and the energy levels in the graphene electrodes, which results in the shift of the self-energy $\tilde\Sigma(\epsilon)=\Sigma(\epsilon-eV_{gl})$.
\begin{figure}[t]
\begin{center}
\epsfxsize=1.\hsize
\epsfbox{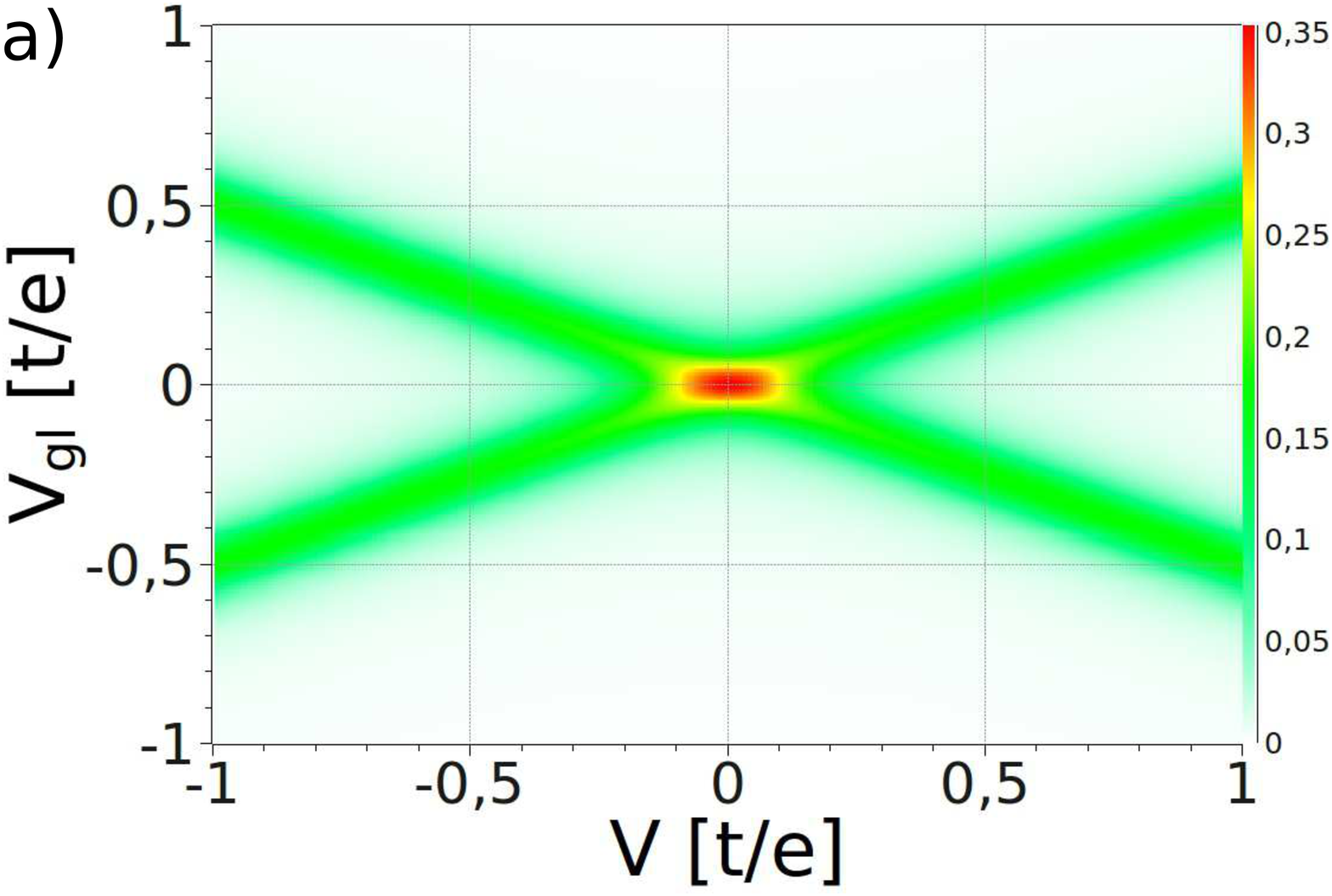}
\epsfxsize=1.\hsize
\epsfbox{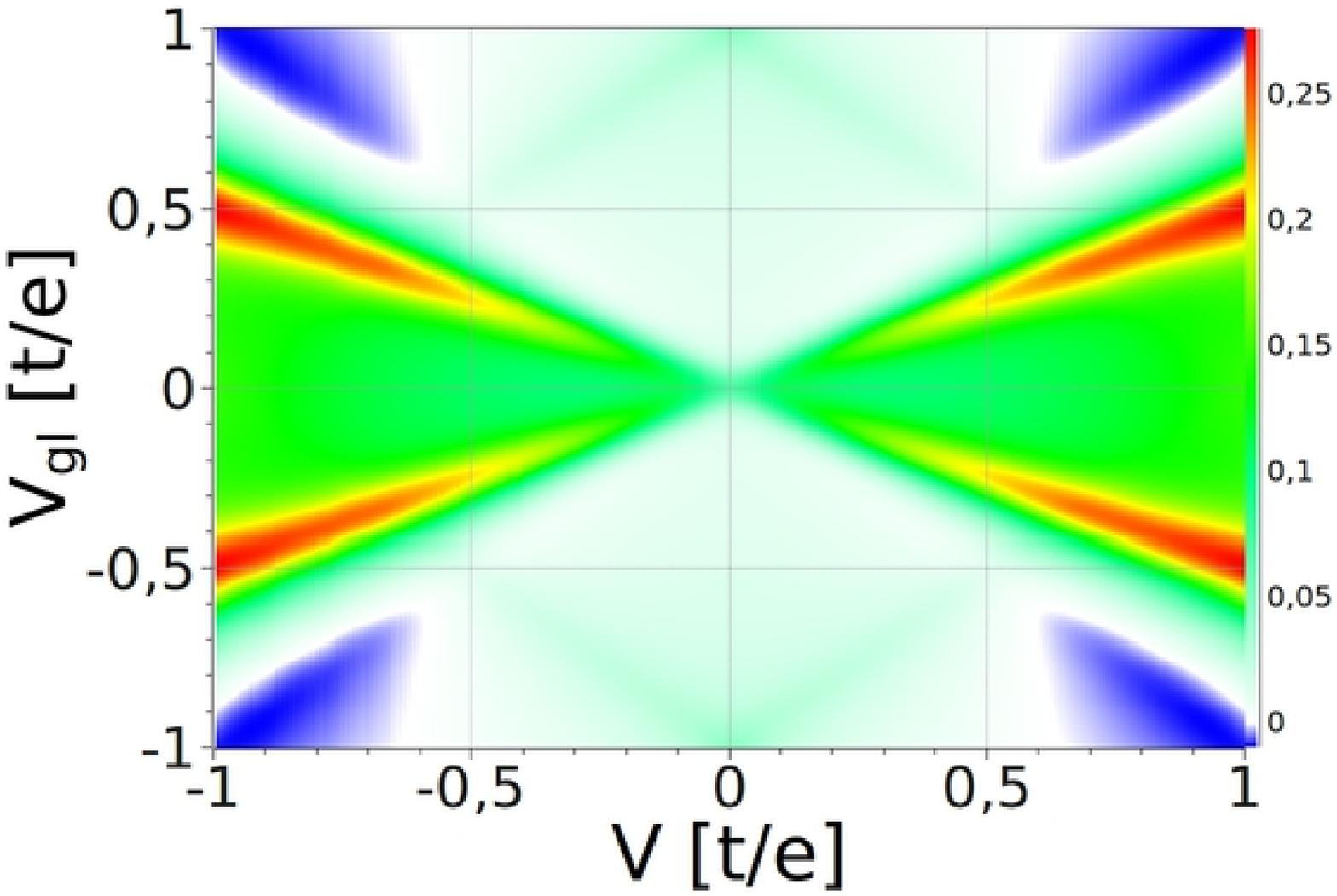}
\epsfxsize=1.\hsize
\epsfbox{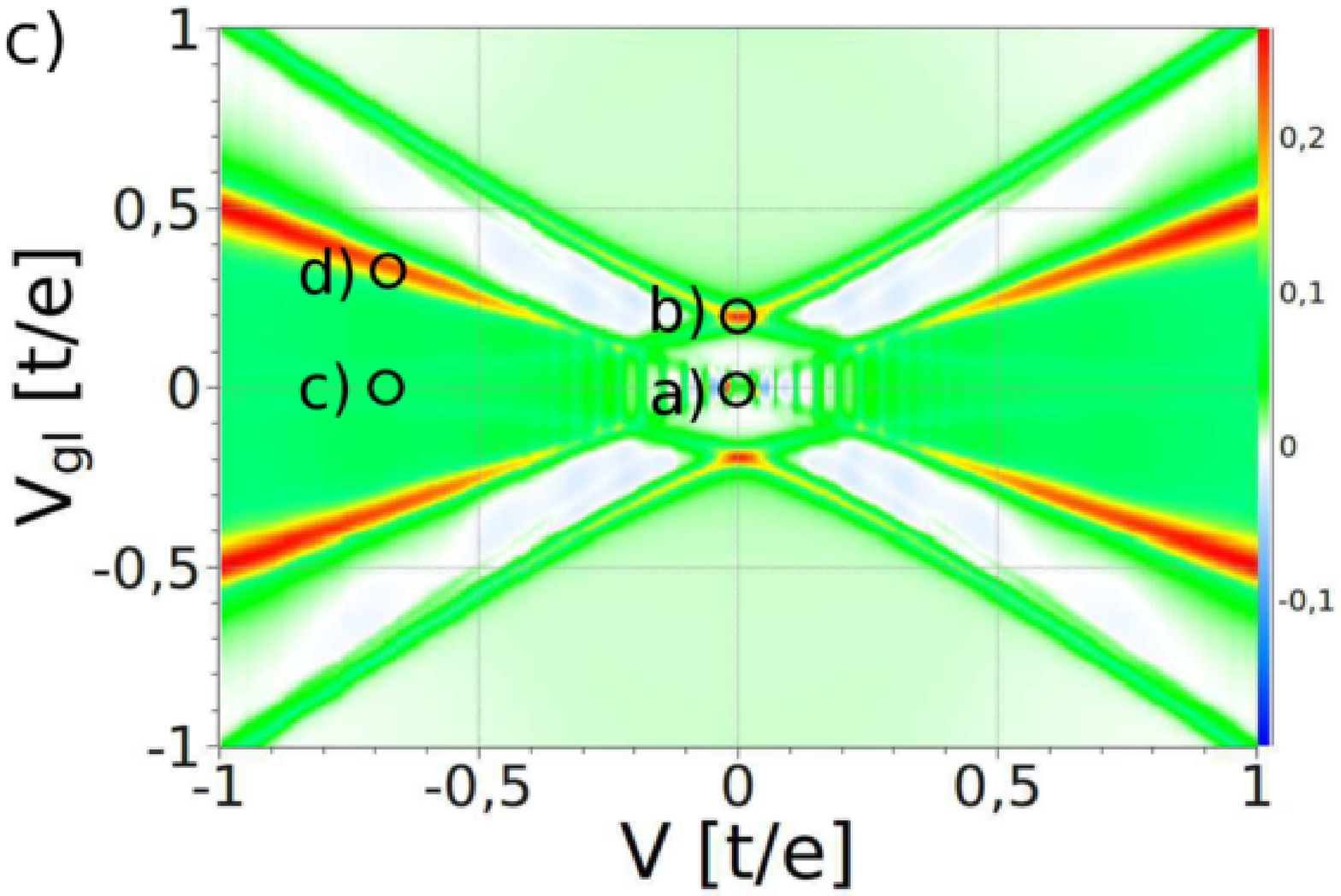}
\caption{(Color online) Differential conductance $dI/dV$ as a function of bias and gate voltage for a single level at $\epsilon_0=0$  coupled to wide-band leads (a), armchair leads (b) and zigzag leads (c), with $V_0=0.1t$ and $k_BT=0.01t$. The scale of the differential conductance is shown at the right.}
\label{dIdV}
\end{center}
\end{figure}

\begin{figure}
\includegraphics[width=0.5\textwidth]{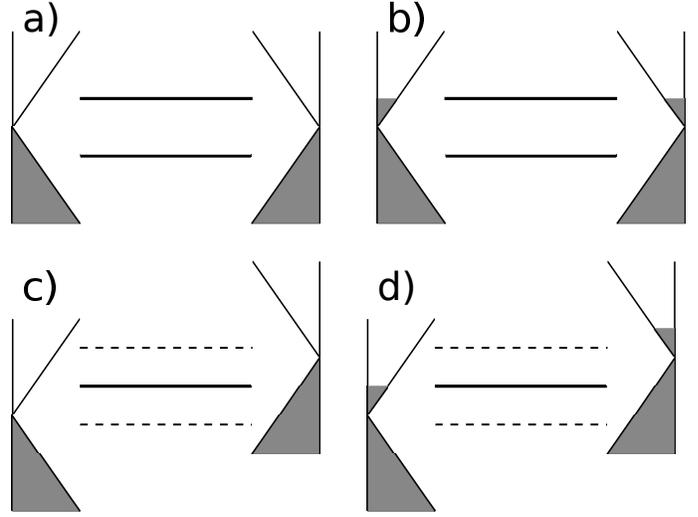}
\caption{The energy level diagrams corresponding to the points a)-d) marked in Fig.\,\ref{dIdV}(c) of the differential conductance for zigzag electrodes. The vertical axis corresponds to the energies. The linear energy dependence of the bulk density of states near the Dirac point is shown schematically for the left and right leads (the singular edge-state density of states at the Dirac point is not shown). See the text for explanations.}
\label{dos}
\end{figure}

In Fig.\,\ref{dIdV} the differential conductance $dI/dV$ for $\epsilon_0=0$ is shown as a function of bias and gate voltage for armchair and zigzag leads in comparison with the wide-band limit. The main features of conductance are explained by the level modification properties, considered previously. For example, at small $V$ and $V_{gl}$ (in the center) the differential conductance has a minimum rather than a maximum, contrary to the metal wide-band case with energy-independent lead self-energy.

In the case of armchair electrodes the level is not shifted and the small conductance at small bias and gate voltage is mainly explained by the small $\Gamma(\epsilon=0)$ that gives also the small current $I\propto\Gamma V$. If we apply the gate voltage, the maximum of differential conductance occurs at $V=\pm 2V_{gl}$, when the Fermi energy crosses the level position, and the current at this point is determined by $\Gamma(\epsilon=V_{gl})$, which is larger at larger $V_{gl}$. On the other hand, the level broadening is also larger in this case. The regions of negative differential conductance at large gate voltage and large bias voltage appear because of the decreasing density of states at $\epsilon>t$.

In the case of zigzag electrodes, this picture is complicated by the edge-state effects, removing levels at small energies and suppressing the transport at small voltages. The additional structures appear due to the existence of the split sublevel. It can lead to interesting effects, when the current at small voltages is determined by the molecular levels with large energies, while the small energy levels are shifted to larger energies. Besides, the level broadening and shift take place for larger $|\epsilon_0|$ similar to the case of the armchair edge. For a further analysis of the $dI/dV$ characteristic in the zigzag case, let us consider the energy level diagrams (Fig.\,\ref{dos}) corresponding to the points a), b), c), d) on Fig.\,\ref{dIdV}c. At point a) the level is split into two equivalent sublevels at finite energy, so that the linear conductance is small. At finite gate voltage (point b)) the Fermi level in the leads is in resonance with one of the sublevels and a distinct local conductance maximum is observed at zero voltage. At zero gate voltage, but finite bias voltage (point c)), the Dirac points and the energies of the edge states are shifted away from zero energy. As a result, the original energy level with $\epsilon_0=0$ is not shifted, but produces two additional side levels at positive and negative energies. The current flows mainly through the unshifted level and is a monotonous function of the voltage, because it is proportional to the density of filled states from one side and the density of empty states from the other side. Similar behavior is observed at this voltages in the armchair case, because edge effects do not play an essential role. Finally, at point d), with $|V|\approx |2V_{gl}|$, the Fermi level in one of the leads crosses the energy level and produces the resonance line. This is similar to the wide-band case, but the current and conductance are larger at larger voltages, because of the energy dependent coupling to the leads.

\section{Conclusions}
\label{sec:conclusions}

In conclusion, we showed that gate effects in graphene molecular junctions are determined by two main effects: the strong energy dependence of the lead self-energies and the shift of these energy-dependent self-energies relative to the Fermi level, together with the usual shift of the molecular levels. In particular, in the zigzag case we demonstrate the splitting of the molecular resonant state. In the biased case with bipolar graphene leads discussed above, removing the nanogap and the molecular level recovers a graphene {\it pn} junction, where the transmission becomes reflectionless at normal incidence. This perfect transmission, decaying with the increase of the incidence angle, is known as Klein tunneling, which has been pioneered for graphene by spin-independent studies~\cite{Katsnelson06naturephysics,Cheianov06prb} and recently generalized to spin-dependent cases.~\cite{Yamakage09epl,Liu12prb} How the picture of Klein tunneling would be modified by the presence of the nanogap with a (single) bridging molecular level is naturally an interesting question that deserves future investigation.

\section*{Acknowledgments}

We thank Michael Thoss and Ivan Pshenichnyuk for fruitful discussions.

This work was funded by the Deutsche Forschungsgemeinschaft within the Collaborative Research Center SFB 689 and the Research Training Group GRK 1570, as well as by the Alexander von Humboldt foundation (M.-H.L.).


\begin{thebibliography}{40}
\expandafter\ifx\csname natexlab\endcsname\relax\def\natexlab#1{#1}\fi
\expandafter\ifx\csname bibnamefont\endcsname\relax
  \def\bibnamefont#1{#1}\fi
\expandafter\ifx\csname bibfnamefont\endcsname\relax
  \def\bibfnamefont#1{#1}\fi
\expandafter\ifx\csname citenamefont\endcsname\relax
  \def\citenamefont#1{#1}\fi
\expandafter\ifx\csname url\endcsname\relax
  \def\url#1{\texttt{#1}}\fi
\expandafter\ifx\csname urlprefix\endcsname\relax\def\urlprefix{URL }\fi
\providecommand{\bibinfo}[2]{#2}
\providecommand{\eprint}[2][]{\url{#2}}

\bibitem[{\citenamefont{Cuniberti et~al.}(2005)\citenamefont{Cuniberti, Fagas,
  and {Richter (Eds.)}}}]{Cuniberti05book}
\bibinfo{author}{\bibfnamefont{G.}~\bibnamefont{Cuniberti}},
  \bibinfo{author}{\bibfnamefont{G.}~\bibnamefont{Fagas}}, \bibnamefont{and}
  \bibinfo{author}{\bibfnamefont{K.}~\bibnamefont{{Richter (Eds.)}}},
  \emph{\bibinfo{title}{Introducing Molecular Electronics}}, vol.
  \bibinfo{volume}{680} of \emph{\bibinfo{series}{Lecture Notes in Physics}}
  (\bibinfo{publisher}{Springer, Berlin}, \bibinfo{year}{2005}).

\bibitem[{\citenamefont{Cuevas and Scheer}(2010)}]{Cuevas10book}
\bibinfo{author}{\bibfnamefont{J.~C.} \bibnamefont{Cuevas}} \bibnamefont{and}
  \bibinfo{author}{\bibfnamefont{E.}~\bibnamefont{Scheer}},
  \emph{\bibinfo{title}{Molecular electronics: An Introduction to Theory and
  Experiment}} (\bibinfo{publisher}{World Scientific}, \bibinfo{year}{2010}).

\bibitem[{\citenamefont{Song et~al.}(2011{\natexlab{a}})\citenamefont{Song,
  Reed, and Lee}}]{Song11advmat}
\bibinfo{author}{\bibfnamefont{H.}~\bibnamefont{Song}},
  \bibinfo{author}{\bibfnamefont{M.~A.} \bibnamefont{Reed}}, \bibnamefont{and}
  \bibinfo{author}{\bibfnamefont{T.}~\bibnamefont{Lee}},
  \bibinfo{journal}{Advanced Materials} \textbf{\bibinfo{volume}{23}},
  \bibinfo{pages}{1583} (\bibinfo{year}{2011}{\natexlab{a}}).

\bibitem[{\citenamefont{Berger et~al.}(2004)\citenamefont{Berger, Song, Li, Li,
  Ogbazghi, Feng, Dai, Marchenkov, Conrad, First et~al.}}]{Berger04jpcb}
\bibinfo{author}{\bibfnamefont{C.}~\bibnamefont{Berger}},
  \bibinfo{author}{\bibfnamefont{Z.}~\bibnamefont{Song}},
  \bibinfo{author}{\bibfnamefont{T.}~\bibnamefont{Li}},
  \bibinfo{author}{\bibfnamefont{X.}~\bibnamefont{Li}},
  \bibinfo{author}{\bibfnamefont{A.~Y.} \bibnamefont{Ogbazghi}},
  \bibinfo{author}{\bibfnamefont{R.}~\bibnamefont{Feng}},
  \bibinfo{author}{\bibfnamefont{Z.}~\bibnamefont{Dai}},
  \bibinfo{author}{\bibfnamefont{A.~N.} \bibnamefont{Marchenkov}},
  \bibinfo{author}{\bibfnamefont{E.~H.} \bibnamefont{Conrad}},
  \bibinfo{author}{\bibfnamefont{P.~N.} \bibnamefont{First}},
  \bibnamefont{et~al.}, \bibinfo{journal}{J. Phys. Chem. B}
  \textbf{\bibinfo{volume}{108}}, \bibinfo{pages}{19912}
  (\bibinfo{year}{2004}).

\bibitem[{\citenamefont{Jia et~al.}(2009)\citenamefont{Jia, Hofmann, Meunier,
  Sumpter, Campos-Delgado, Romo-Herrera, Son, Hsieh, Reina, Kong
  et~al.}}]{Jia09science}
\bibinfo{author}{\bibfnamefont{X.}~\bibnamefont{Jia}},
  \bibinfo{author}{\bibfnamefont{M.}~\bibnamefont{Hofmann}},
  \bibinfo{author}{\bibfnamefont{V.}~\bibnamefont{Meunier}},
  \bibinfo{author}{\bibfnamefont{B.~G.} \bibnamefont{Sumpter}},
  \bibinfo{author}{\bibfnamefont{J.}~\bibnamefont{Campos-Delgado}},
  \bibinfo{author}{\bibfnamefont{J.~M.} \bibnamefont{Romo-Herrera}},
  \bibinfo{author}{\bibfnamefont{H.}~\bibnamefont{Son}},
  \bibinfo{author}{\bibfnamefont{Y.-P.} \bibnamefont{Hsieh}},
  \bibinfo{author}{\bibfnamefont{A.}~\bibnamefont{Reina}},
  \bibinfo{author}{\bibfnamefont{J.}~\bibnamefont{Kong}}, \bibnamefont{et~al.},
  \bibinfo{journal}{Science} \textbf{\bibinfo{volume}{323}},
  \bibinfo{pages}{1701} (\bibinfo{year}{2009}).

\bibitem[{\citenamefont{Jin et~al.}(2009)\citenamefont{Jin, Lan, Peng, Suenaga,
  and Iijima}}]{Jin09prl}
\bibinfo{author}{\bibfnamefont{C.}~\bibnamefont{Jin}},
  \bibinfo{author}{\bibfnamefont{H.}~\bibnamefont{Lan}},
  \bibinfo{author}{\bibfnamefont{L.}~\bibnamefont{Peng}},
  \bibinfo{author}{\bibfnamefont{K.}~\bibnamefont{Suenaga}}, \bibnamefont{and}
  \bibinfo{author}{\bibfnamefont{S.}~\bibnamefont{Iijima}},
  \bibinfo{journal}{Phys. Rev. Lett.} \textbf{\bibinfo{volume}{102}},
  \bibinfo{pages}{205501} (\bibinfo{year}{2009}).

\bibitem[{\citenamefont{Chuvilin et~al.}(2009)\citenamefont{Chuvilin, Meyer,
  Algara-Siller, and Kaiser}}]{Chuvilin09njp}
\bibinfo{author}{\bibfnamefont{A.}~\bibnamefont{Chuvilin}},
  \bibinfo{author}{\bibfnamefont{J.~C.} \bibnamefont{Meyer}},
  \bibinfo{author}{\bibfnamefont{G.}~\bibnamefont{Algara-Siller}},
  \bibnamefont{and} \bibinfo{author}{\bibfnamefont{U.}~\bibnamefont{Kaiser}},
  \bibinfo{journal}{New J. Phys.} \textbf{\bibinfo{volume}{11}},
  \bibinfo{pages}{083019} (\bibinfo{year}{2009}).

\bibitem[{\citenamefont{He et~al.}(2010)\citenamefont{He, Dong, Li, Wang, Shao,
  Zhang, Jiang, and Hu}}]{He10apl}
\bibinfo{author}{\bibfnamefont{Y.}~\bibnamefont{He}},
  \bibinfo{author}{\bibfnamefont{H.}~\bibnamefont{Dong}},
  \bibinfo{author}{\bibfnamefont{T.}~\bibnamefont{Li}},
  \bibinfo{author}{\bibfnamefont{C.}~\bibnamefont{Wang}},
  \bibinfo{author}{\bibfnamefont{W.}~\bibnamefont{Shao}},
  \bibinfo{author}{\bibfnamefont{Y.}~\bibnamefont{Zhang}},
  \bibinfo{author}{\bibfnamefont{L.}~\bibnamefont{Jiang}}, \bibnamefont{and}
  \bibinfo{author}{\bibfnamefont{W.}~\bibnamefont{Hu}}, \bibinfo{journal}{Appl.
  Phys. Lett.} \textbf{\bibinfo{volume}{97}}, \bibinfo{eid}{133301}
  (\bibinfo{year}{2010}).

\bibitem[{\citenamefont{Song et~al.}(2011{\natexlab{b}})\citenamefont{Song,
  Schneider, Xu, Pandraud, Dekker, and Zandbergen}}]{Song11nanolett}
\bibinfo{author}{\bibfnamefont{B.}~\bibnamefont{Song}},
  \bibinfo{author}{\bibfnamefont{G.~F.} \bibnamefont{Schneider}},
  \bibinfo{author}{\bibfnamefont{Q.}~\bibnamefont{Xu}},
  \bibinfo{author}{\bibfnamefont{G.}~\bibnamefont{Pandraud}},
  \bibinfo{author}{\bibfnamefont{C.}~\bibnamefont{Dekker}}, \bibnamefont{and}
  \bibinfo{author}{\bibfnamefont{H.}~\bibnamefont{Zandbergen}},
  \bibinfo{journal}{Nano Lett.} \textbf{\bibinfo{volume}{11}},
  \bibinfo{pages}{2247} (\bibinfo{year}{2011}{\natexlab{b}}).

\bibitem[{\citenamefont{Gutierrez et~al.}(2002)\citenamefont{Gutierrez, Fagas,
  Cuniberti, Grossmann, Schmidt, and Richter}}]{Gutierrez02prb}
\bibinfo{author}{\bibfnamefont{R.}~\bibnamefont{Gutierrez}},
  \bibinfo{author}{\bibfnamefont{G.}~\bibnamefont{Fagas}},
  \bibinfo{author}{\bibfnamefont{G.}~\bibnamefont{Cuniberti}},
  \bibinfo{author}{\bibfnamefont{F.}~\bibnamefont{Grossmann}},
  \bibinfo{author}{\bibfnamefont{R.}~\bibnamefont{Schmidt}}, \bibnamefont{and}
  \bibinfo{author}{\bibfnamefont{K.}~\bibnamefont{Richter}},
  \bibinfo{journal}{Phys. Rev. B} \textbf{\bibinfo{volume}{65}},
  \bibinfo{pages}{113410} (\bibinfo{year}{2002}).

\bibitem[{\citenamefont{Chen et~al.}(2007)\citenamefont{Chen, Zhang, and
  Hybertsen}}]{Chen07prb}
\bibinfo{author}{\bibfnamefont{Y.-R.} \bibnamefont{Chen}},
  \bibinfo{author}{\bibfnamefont{L.}~\bibnamefont{Zhang}}, \bibnamefont{and}
  \bibinfo{author}{\bibfnamefont{M.~S.} \bibnamefont{Hybertsen}},
  \bibinfo{journal}{Phys. Rev. B} \textbf{\bibinfo{volume}{76}},
  \bibinfo{pages}{115408} (\bibinfo{year}{2007}).

\bibitem[{\citenamefont{Fagas et~al.}(2004)\citenamefont{Fagas, Kambili, and
  Elstner}}]{Fagas04cpl}
\bibinfo{author}{\bibfnamefont{G.}~\bibnamefont{Fagas}},
  \bibinfo{author}{\bibfnamefont{A.}~\bibnamefont{Kambili}}, \bibnamefont{and}
  \bibinfo{author}{\bibfnamefont{M.}~\bibnamefont{Elstner}},
  \bibinfo{journal}{Chem. Phys. Lett.} \textbf{\bibinfo{volume}{389}},
  \bibinfo{pages}{268 } (\bibinfo{year}{2004}).

\bibitem[{\citenamefont{Fagas and Kambili}(2004)}]{Fagas0403694}
\bibinfo{author}{\bibfnamefont{G.}~\bibnamefont{Fagas}} \bibnamefont{and}
  \bibinfo{author}{\bibfnamefont{A.}~\bibnamefont{Kambili}},
  \bibinfo{journal}{arXiv:cond-mat/0403694}  (\bibinfo{year}{2004}).

\bibitem[{\citenamefont{Cheraghchi and Esfarjani}(2008)}]{Cheraghchi08prb}
\bibinfo{author}{\bibfnamefont{H.}~\bibnamefont{Cheraghchi}} \bibnamefont{and}
  \bibinfo{author}{\bibfnamefont{K.}~\bibnamefont{Esfarjani}},
  \bibinfo{journal}{Phys. Rev. B} \textbf{\bibinfo{volume}{78}},
  \bibinfo{pages}{085123} (\bibinfo{year}{2008}).

\bibitem[{\citenamefont{Yin et~al.}(2009)\citenamefont{Yin, Liang, Jiang, Chen,
  Wang, Note, Mizuseki, and Kawazoe}}]{Yin09jcp}
\bibinfo{author}{\bibfnamefont{G.}~\bibnamefont{Yin}},
  \bibinfo{author}{\bibfnamefont{Y.~Y.} \bibnamefont{Liang}},
  \bibinfo{author}{\bibfnamefont{F.}~\bibnamefont{Jiang}},
  \bibinfo{author}{\bibfnamefont{H.}~\bibnamefont{Chen}},
  \bibinfo{author}{\bibfnamefont{P.}~\bibnamefont{Wang}},
  \bibinfo{author}{\bibfnamefont{R.}~\bibnamefont{Note}},
  \bibinfo{author}{\bibfnamefont{H.}~\bibnamefont{Mizuseki}}, \bibnamefont{and}
  \bibinfo{author}{\bibfnamefont{Y.}~\bibnamefont{Kawazoe}},
  \bibinfo{journal}{J. Chem. Phys.} \textbf{\bibinfo{volume}{131}},
  \bibinfo{eid}{234706} (\bibinfo{year}{2009}).

\bibitem[{\citenamefont{Chen et~al.}(2009)\citenamefont{Chen, Andreev, and
  Bertsch}}]{Chen09prb}
\bibinfo{author}{\bibfnamefont{W.}~\bibnamefont{Chen}},
  \bibinfo{author}{\bibfnamefont{A.~V.} \bibnamefont{Andreev}},
  \bibnamefont{and} \bibinfo{author}{\bibfnamefont{G.~F.}
  \bibnamefont{Bertsch}}, \bibinfo{journal}{Phys. Rev. B}
  \textbf{\bibinfo{volume}{80}}, \bibinfo{pages}{085410}
  (\bibinfo{year}{2009}).

\bibitem[{\citenamefont{F\"urst et~al.}(2010)\citenamefont{F\"urst, Brandbyge,
  and Jauho}}]{Fuerstl10epl}
\bibinfo{author}{\bibfnamefont{J.~A.} \bibnamefont{F\"urst}},
  \bibinfo{author}{\bibfnamefont{M.}~\bibnamefont{Brandbyge}},
  \bibnamefont{and} \bibinfo{author}{\bibfnamefont{A.-P.} \bibnamefont{Jauho}},
  \bibinfo{journal}{EPL} \textbf{\bibinfo{volume}{91}}, \bibinfo{pages}{37002}
  (\bibinfo{year}{2010}).

\bibitem[{\citenamefont{Shen et~al.}(2010)\citenamefont{Shen, Zeng, Yang,
  Zhang, Wang, and Feng}}]{Shen10jacs}
\bibinfo{author}{\bibfnamefont{L.}~\bibnamefont{Shen}},
  \bibinfo{author}{\bibfnamefont{M.}~\bibnamefont{Zeng}},
  \bibinfo{author}{\bibfnamefont{S.-W.} \bibnamefont{Yang}},
  \bibinfo{author}{\bibfnamefont{C.}~\bibnamefont{Zhang}},
  \bibinfo{author}{\bibfnamefont{X.}~\bibnamefont{Wang}}, \bibnamefont{and}
  \bibinfo{author}{\bibfnamefont{Y.}~\bibnamefont{Feng}}, \bibinfo{journal}{J.
  Am. Chem. Soc.} \textbf{\bibinfo{volume}{132}}, \bibinfo{pages}{11481}
  (\bibinfo{year}{2010}).

\bibitem[{\citenamefont{Saha et~al.}(2010)\citenamefont{Saha,
  Nikoli\ifmmode~\acute{c}\else \'{c}\fi{}, Meunier, Lu, and
  Bernholc}}]{Saha10prl}
\bibinfo{author}{\bibfnamefont{K.~K.} \bibnamefont{Saha}},
  \bibinfo{author}{\bibfnamefont{B.~K.}
  \bibnamefont{Nikoli\ifmmode~\acute{c}\else \'{c}\fi{}}},
  \bibinfo{author}{\bibfnamefont{V.}~\bibnamefont{Meunier}},
  \bibinfo{author}{\bibfnamefont{W.}~\bibnamefont{Lu}}, \bibnamefont{and}
  \bibinfo{author}{\bibfnamefont{J.}~\bibnamefont{Bernholc}},
  \bibinfo{journal}{Phys. Rev. Lett.} \textbf{\bibinfo{volume}{105}},
  \bibinfo{pages}{236803} (\bibinfo{year}{2010}).

\bibitem[{\citenamefont{Kawai et~al.}(2011)\citenamefont{Kawai, Poetschke,
  Miyamoto, Rocha, Roche, and Cuniberti}}]{Kawai11prb}
\bibinfo{author}{\bibfnamefont{T.}~\bibnamefont{Kawai}},
  \bibinfo{author}{\bibfnamefont{M.}~\bibnamefont{Poetschke}},
  \bibinfo{author}{\bibfnamefont{Y.}~\bibnamefont{Miyamoto}},
  \bibinfo{author}{\bibfnamefont{C.~G.} \bibnamefont{Rocha}},
  \bibinfo{author}{\bibfnamefont{S.}~\bibnamefont{Roche}}, \bibnamefont{and}
  \bibinfo{author}{\bibfnamefont{G.}~\bibnamefont{Cuniberti}},
  \bibinfo{journal}{Phys. Rev. B} \textbf{\bibinfo{volume}{83}},
  \bibinfo{pages}{241405} (\bibinfo{year}{2011}).

\bibitem[{\citenamefont{Agapito and Cheng}(2007)}]{Agapito07jpcc}
\bibinfo{author}{\bibfnamefont{L.~A.} \bibnamefont{Agapito}} \bibnamefont{and}
  \bibinfo{author}{\bibfnamefont{H.-P.} \bibnamefont{Cheng}},
  \bibinfo{journal}{J. Phys. Chem. C} \textbf{\bibinfo{volume}{111}},
  \bibinfo{pages}{14266} (\bibinfo{year}{2007}).

\bibitem[{\citenamefont{Motta et~al.}(2011)\citenamefont{Motta, Trioni, Brivio,
  and Sebastian}}]{Motta11prb}
\bibinfo{author}{\bibfnamefont{C.}~\bibnamefont{Motta}},
  \bibinfo{author}{\bibfnamefont{M.~I.} \bibnamefont{Trioni}},
  \bibinfo{author}{\bibfnamefont{G.~P.} \bibnamefont{Brivio}},
  \bibnamefont{and} \bibinfo{author}{\bibfnamefont{K.~L.}
  \bibnamefont{Sebastian}}, \bibinfo{journal}{Phys. Rev. B}
  \textbf{\bibinfo{volume}{84}}, \bibinfo{pages}{113408}
  (\bibinfo{year}{2011}).

\bibitem[{\citenamefont{Cai et~al.}(2011)\citenamefont{Cai, Zhang, Zhang, and
  Feng}}]{Cai11eprint}
\bibinfo{author}{\bibfnamefont{Y.}~\bibnamefont{Cai}},
  \bibinfo{author}{\bibfnamefont{A.}~\bibnamefont{Zhang}},
  \bibinfo{author}{\bibfnamefont{C.}~\bibnamefont{Zhang}}, \bibnamefont{and}
  \bibinfo{author}{\bibfnamefont{Y.~P.} \bibnamefont{Feng}},
  \bibinfo{journal}{arXiv:1111.1811}  (\bibinfo{year}{2011}).

\bibitem[{\citenamefont{Carrascal et~al.}(2012)\citenamefont{Carrascal,
  Garc\'ia-Su\'arez, and Ferrer}}]{Carrascal12eprint}
\bibinfo{author}{\bibfnamefont{D.}~\bibnamefont{Carrascal}},
  \bibinfo{author}{\bibfnamefont{V.~M.} \bibnamefont{Garc\'ia-Su\'arez}},
  \bibnamefont{and} \bibinfo{author}{\bibfnamefont{J.}~\bibnamefont{Ferrer}},
  \bibinfo{journal}{arXiv:1202.2699}  (\bibinfo{year}{2012}).

\bibitem[{\citenamefont{Prins et~al.}(2011)\citenamefont{Prins, Barreiro,
  Ruitenberg, Seldenthuis, Aliaga-Alcalde, Vandersypen, and van~der
  Zant}}]{Prins11nanolett}
\bibinfo{author}{\bibfnamefont{F.}~\bibnamefont{Prins}},
  \bibinfo{author}{\bibfnamefont{A.}~\bibnamefont{Barreiro}},
  \bibinfo{author}{\bibfnamefont{J.~W.} \bibnamefont{Ruitenberg}},
  \bibinfo{author}{\bibfnamefont{J.~S.} \bibnamefont{Seldenthuis}},
  \bibinfo{author}{\bibfnamefont{N.}~\bibnamefont{Aliaga-Alcalde}},
  \bibinfo{author}{\bibfnamefont{L.~M.~K.} \bibnamefont{Vandersypen}},
  \bibnamefont{and} \bibinfo{author}{\bibfnamefont{H.~S.~J.}
  \bibnamefont{van~der Zant}}, \bibinfo{journal}{Nano Lett.}
  \textbf{\bibinfo{volume}{11}}, \bibinfo{pages}{4607} (\bibinfo{year}{2011}).

\bibitem[{\citenamefont{Wallace}(1947)}]{Wallace47prb}
\bibinfo{author}{\bibfnamefont{P.~R.} \bibnamefont{Wallace}},
  \bibinfo{journal}{Phys. Rev.} \textbf{\bibinfo{volume}{71}},
  \bibinfo{pages}{622} (\bibinfo{year}{1947}).

\bibitem[{\citenamefont{Nakada et~al.}(1996)\citenamefont{Nakada, Fujita,
  Dresselhaus, and Dresselhaus}}]{Nakada96prb}
\bibinfo{author}{\bibfnamefont{K.}~\bibnamefont{Nakada}},
  \bibinfo{author}{\bibfnamefont{M.}~\bibnamefont{Fujita}},
  \bibinfo{author}{\bibfnamefont{G.}~\bibnamefont{Dresselhaus}},
  \bibnamefont{and} \bibinfo{author}{\bibfnamefont{M.~S.}
  \bibnamefont{Dresselhaus}}, \bibinfo{journal}{Phys. Rev. B}
  \textbf{\bibinfo{volume}{54}}, \bibinfo{pages}{17954} (\bibinfo{year}{1996}).

\bibitem[{\citenamefont{Fujita et~al.}(1996)\citenamefont{Fujita, Wakabayashi,
  Nakada, and Kusakabe}}]{Fujita96jpsj}
\bibinfo{author}{\bibfnamefont{M.}~\bibnamefont{Fujita}},
  \bibinfo{author}{\bibfnamefont{K.}~\bibnamefont{Wakabayashi}},
  \bibinfo{author}{\bibfnamefont{K.}~\bibnamefont{Nakada}}, \bibnamefont{and}
  \bibinfo{author}{\bibfnamefont{K.}~\bibnamefont{Kusakabe}},
  \bibinfo{journal}{J. Phys. Soc.Jap.} \textbf{\bibinfo{volume}{65}},
  \bibinfo{pages}{1920} (\bibinfo{year}{1996}).

\bibitem[{\citenamefont{Niimi et~al.}(2006)\citenamefont{Niimi, Matsui,
  Kambara, Tagami, Tsukada, and Fukuyama}}]{Niimi06prb}
\bibinfo{author}{\bibfnamefont{Y.}~\bibnamefont{Niimi}},
  \bibinfo{author}{\bibfnamefont{T.}~\bibnamefont{Matsui}},
  \bibinfo{author}{\bibfnamefont{H.}~\bibnamefont{Kambara}},
  \bibinfo{author}{\bibfnamefont{K.}~\bibnamefont{Tagami}},
  \bibinfo{author}{\bibfnamefont{M.}~\bibnamefont{Tsukada}}, \bibnamefont{and}
  \bibinfo{author}{\bibfnamefont{H.}~\bibnamefont{Fukuyama}},
  \bibinfo{journal}{Phys. Rev. B} \textbf{\bibinfo{volume}{73}},
  \bibinfo{pages}{085421} (\bibinfo{year}{2006}).

\bibitem[{\citenamefont{Kobayashi et~al.}(2006)\citenamefont{Kobayashi, Fukui,
  Enoki, and Kusakabe}}]{Kobayashi06prb}
\bibinfo{author}{\bibfnamefont{Y.}~\bibnamefont{Kobayashi}},
  \bibinfo{author}{\bibfnamefont{K.-i.} \bibnamefont{Fukui}},
  \bibinfo{author}{\bibfnamefont{T.}~\bibnamefont{Enoki}}, \bibnamefont{and}
  \bibinfo{author}{\bibfnamefont{K.}~\bibnamefont{Kusakabe}},
  \bibinfo{journal}{Phys. Rev. B} \textbf{\bibinfo{volume}{73}},
  \bibinfo{pages}{125415} (\bibinfo{year}{2006}).

\bibitem[{\citenamefont{{Lopez~Sancho}
  et~al.}(1985{\natexlab{a}})\citenamefont{{Lopez~Sancho}, {Lopez~Sancho}, and
  Rubio}}]{LopezSancho84jpf}
\bibinfo{author}{\bibfnamefont{M.~P.} \bibnamefont{{Lopez~Sancho}}},
  \bibinfo{author}{\bibfnamefont{J.~M.} \bibnamefont{{Lopez~Sancho}}},
  \bibnamefont{and} \bibinfo{author}{\bibfnamefont{J.}~\bibnamefont{Rubio}},
  \bibinfo{journal}{J. Phys. F: Met. Phys.} \textbf{\bibinfo{volume}{14}},
  \bibinfo{pages}{1205} (\bibinfo{year}{1985}{\natexlab{a}}).

\bibitem[{\citenamefont{{Lopez~Sancho}
  et~al.}(1985{\natexlab{b}})\citenamefont{{Lopez~Sancho}, {Lopez~Sancho}, and
  Rubio}}]{LopezSancho85jpf}
\bibinfo{author}{\bibfnamefont{M.~P.} \bibnamefont{{Lopez~Sancho}}},
  \bibinfo{author}{\bibfnamefont{J.~M.} \bibnamefont{{Lopez~Sancho}}},
  \bibnamefont{and} \bibinfo{author}{\bibfnamefont{J.}~\bibnamefont{Rubio}},
  \bibinfo{journal}{J. Phys. F: Met. Phys.} \textbf{\bibinfo{volume}{15}},
  \bibinfo{pages}{851} (\bibinfo{year}{1985}{\natexlab{b}}).

\bibitem[{\citenamefont{Wimmer}(2008)}]{Wimmer08thesis}
\bibinfo{author}{\bibfnamefont{M.}~\bibnamefont{Wimmer}}, \bibinfo{type}{New},
  \bibinfo{school}{University of Regensburg} (\bibinfo{year}{2008}).

\bibitem[{\citenamefont{Meir and Wingreen}(1992)}]{Meir92prl}
\bibinfo{author}{\bibfnamefont{Y.}~\bibnamefont{Meir}} \bibnamefont{and}
  \bibinfo{author}{\bibfnamefont{N.~S.} \bibnamefont{Wingreen}},
  \bibinfo{journal}{Phys. Rev. Lett.} \textbf{\bibinfo{volume}{68}},
  \bibinfo{pages}{2512} (\bibinfo{year}{1992}).

\bibitem[{\citenamefont{Jauho et~al.}(1994)\citenamefont{Jauho, Wingreen, and
  Meir}}]{Jauho94prb}
\bibinfo{author}{\bibfnamefont{A.-P.} \bibnamefont{Jauho}},
  \bibinfo{author}{\bibfnamefont{N.~S.} \bibnamefont{Wingreen}},
  \bibnamefont{and} \bibinfo{author}{\bibfnamefont{Y.}~\bibnamefont{Meir}},
  \bibinfo{journal}{Phys. Rev. B} \textbf{\bibinfo{volume}{50}},
  \bibinfo{pages}{5528} (\bibinfo{year}{1994}).

\bibitem[{\citenamefont{Jauho}(2006)}]{Jauho06jpcs}
\bibinfo{author}{\bibfnamefont{A.-P.} \bibnamefont{Jauho}},
  \bibinfo{journal}{Journal of Physics: Conference Series}
  \textbf{\bibinfo{volume}{35}}, \bibinfo{pages}{313} (\bibinfo{year}{2006}).

\bibitem[{\citenamefont{Katsnelson et~al.}(2006)\citenamefont{Katsnelson,
  Novoselov, and Geim}}]{Katsnelson06naturephysics}
\bibinfo{author}{\bibfnamefont{M.~I.} \bibnamefont{Katsnelson}},
  \bibinfo{author}{\bibfnamefont{K.~S.} \bibnamefont{Novoselov}},
  \bibnamefont{and} \bibinfo{author}{\bibfnamefont{A.~K.} \bibnamefont{Geim}},
  \bibinfo{journal}{Nature Physics} \textbf{\bibinfo{volume}{2}},
  \bibinfo{pages}{620} (\bibinfo{year}{2006}).

\bibitem[{\citenamefont{Cheianov and Fal'ko}(2006)}]{Cheianov06prb}
\bibinfo{author}{\bibfnamefont{V.~V.} \bibnamefont{Cheianov}} \bibnamefont{and}
  \bibinfo{author}{\bibfnamefont{V.~I.} \bibnamefont{Fal'ko}},
  \bibinfo{journal}{Phys. Rev. B} \textbf{\bibinfo{volume}{74}},
  \bibinfo{pages}{041403} (\bibinfo{year}{2006}).

\bibitem[{\citenamefont{Yamakage et~al.}(2009)\citenamefont{Yamakage, Imura,
  Cayssol, and Kuramoto}}]{Yamakage09epl}
\bibinfo{author}{\bibfnamefont{A.}~\bibnamefont{Yamakage}},
  \bibinfo{author}{\bibfnamefont{K.~I.} \bibnamefont{Imura}},
  \bibinfo{author}{\bibfnamefont{J.}~\bibnamefont{Cayssol}}, \bibnamefont{and}
  \bibinfo{author}{\bibfnamefont{Y.}~\bibnamefont{Kuramoto}},
  \bibinfo{journal}{EPL} \textbf{\bibinfo{volume}{87}}, \bibinfo{pages}{47005}
  (\bibinfo{year}{2009}).

\bibitem[{\citenamefont{Liu et~al.}(2012)\citenamefont{Liu, Bundesmann, and
  Richter}}]{Liu12prb}
\bibinfo{author}{\bibfnamefont{M.-H.} \bibnamefont{Liu}},
  \bibinfo{author}{\bibfnamefont{J.}~\bibnamefont{Bundesmann}},
  \bibnamefont{and} \bibinfo{author}{\bibfnamefont{K.}~\bibnamefont{Richter}},
  \bibinfo{journal}{Phys. Rev. B} \textbf{\bibinfo{volume}{85}},
  \bibinfo{pages}{085406} (\bibinfo{year}{2012}).

\end{thebibliography}
\end{document}